\documentclass[longauth]{aa} 

\usepackage{graphicx}
\usepackage[varg]{txfonts}

\usepackage{hyperref}
\hypersetup{colorlinks=true, urlcolor=blue, citecolor=blue}

\usepackage{color}
\usepackage{ulem}

\usepackage{textcomp}

\begin{document}

\title{Characterization of the MASCOT landing area by Hayabusa2}

\author{Stefan~Schröder \inst{1}\fnmsep\thanks{Corresponding author, \email{stefanus.schroder@ltu.se}} \and Naoya~Sakatani \inst{2} \and Rie~Honda \inst{3} \and Eri~Tatsumi \inst{5} \and Yasuhiro~Yokota \inst{2,4} \and Deborah~Domingue \inst{6} \and Yuichiro~Cho \inst{7} \and Shingo~Kameda \inst{8} \and Kohei~Kitazato \inst{9} \and Toru~Kouyama \inst{10} \and Moe~Matsuoka \inst{11} \and Akira Miura \inst{2} \and Tomokatsu~Morota \inst{7} \and Tatsuaki Okada \inst{2,7} \and Hirotaka~Sawada \inst{2} \and Hiroki Senshu \inst{12} \and Yuri Shimaki \inst{2} \and Seiji~Sugita \inst{7} \and Satoshi~Tanaka \inst{2,14,15} \and Hikaru~Yabuta \inst{16} \and Manabu~Yamada \inst{12} \and Matthias~Grott \inst{13} \and Maximilian Hamm \inst{13,17} \and Tra-Mi~Ho \inst{18} \and Ralf~Jaumann \inst{19} \and Stefano~Mottola \inst{13} \and Katharina~Otto \inst{13} \and Nicole~Schmitz \inst{13} \and Frank~Scholten \inst{13}}

\institute{Luleå University of Technology, 98128 Kiruna, Sweden \and Institute of Space and Astronautical Science, Japan Aerospace Exploration Agency, Sagamihara, Kanagawa 252-5210, Japan \and Center for Data Science, Ehime University, Matsuyama, Ehime 790–8577, Japan \and Kochi University, Kochi 780-8520, Japan \and Instituto de Astrofísica de Canarias, University of La Laguna, 38205 La Laguna, Tenerife, Spain \and Planetary Science Institute, Tucson, AZ 85719-2395, USA \and University of Tokyo, Bunkyo, Tokyo 113-0033, Japan \and Rikkyo University, Toshima, Tokyo 171-8501, Japan \and University of Aizu, Fukushima 965-8580, Japan \and National Institute of Advanced Industrial Science and Technology, Koto, Tokyo 135-0064, Japan \and National Institute of Advanced Industrial Science and Technology, Tsukuba, Ibaraki 305-8567, Japan \and Planetary Exploration Research Center, Chiba Institute of Technology, Narashino, Chiba 275-0016, Japan \and Institute of Planetary Research, German Aerospace Center (DLR), 12489 Berlin, Germany \and Graduate University for Advanced Studies, SOKENDAI, Hayama, Kanagawa 240-0193, Japan \and University of Tokyo, Kashiwa, Chiba 277-8561, Japan \and Hiroshima University, 739-8526 Hiroshima, Japan \and University of Potsdam, 14469 Potsdam, Germany \and Institute of Space Systems, German Aerospace Center (DLR), 28359 Bremen, Germany \and Institute of Geological Sciences, Free University of Berlin, 12249 Berlin, Germany}

\abstract
{After landing on C-type asteroid Ryugu, MASCOT imaged brightly colored, submillimeter-sized inclusions in a small rock. Hayabusa2 successfully returned a sample of small particles from the surface of Ryugu, but none of these appear to harbor such inclusions. The samples are considered representative of Ryugu.}
{To understand the apparent discrepancy between MASCOT observations and Ryugu samples, we assess whether the MASCOT landing site, and the rock by implication, is perhaps atypical for Ryugu.}
{We analyzed observations of the MASCOT landing area acquired by three instruments on board Hayabusa2: a camera (ONC), a near-infrared spectrometer (NIRS3), and a thermal infrared imager (TIR). We compared the landing area properties thus retrieved with those of the average Ryugu surface.}
{We selected several areas and landforms in the landing area for analysis: a small crater, a collection of smooth rocks, and the landing site itself. The crater is relatively blue and the rocks are relatively red. The spectral and thermophysical properties of the landing site are very close to those of the average Ryugu surface. The spectral properties of the MASCOT rock are probably close to average, but its thermal inertia may be somewhat higher.}
{The MASCOT rock can also be considered representative of Ryugu. Some of the submillimeter-sized particles in the returned samples stand out because of their atypical spectral properties. Such particles may be present as inclusions in the MASCOT rock.}

\keywords{Methods: data analysis -- Minor planets, asteroids: individual: Ryugu
}

\maketitle

\section{Introduction} \label{sec:intro}

On 3~October 2018, a lander was released by the Hayabusa2 spacecraft of the Japan Aerospace Exploration Agency (JAXA) toward the surface of 162173~Ryugu, a small near-Earth asteroid of the taxonomic C type. The Mobile Asteroid surface SCOuT (MASCOT) was a small package equipped with a camera, spectrometer, radiometer, and magnetometer \citep{H17}. The little lander successfully completed its 17-hour-long mission at the asteroid, during which all instruments collected data \citep{Ho20}. The onboard MASCam camera acquired 20~images while descending to the surface, of which roughly half show the area around the landing site \citep{J19}. After landing, MASCOT settled in front of a rock that was a bit larger than the lander itself. Multispectral imaging of the rock revealed a dark matrix with numerous small, relatively bright, multicolored inclusions \citep{J19,S20}. While the inclusions appear similar to those in carbonaceous chondrite (CC) meteorites, the rock is not easily associated with any particular CC group \citep{S20,O21b}. The rock’s absolute reflectance as measured by MASCOT is consistent with Ryugu’s average reflectance \citep{S20}. The MASCOT radiometer (MARA) found the rock to possess an unusually low thermal inertia, which suggests a high porosity \citep{G19,H22}.

Whereas MASCOT retrieved the detailed properties of a single Ryugu rock, instruments on board Hayabusa2 mapped the entire asteroid surface multiple times over the course of its mission. To the optical navigation camera (ONC), the surface of Ryugu appeared dark and generally uniform in color, with abundant rocks and boulders \citep{Su19}. Small boulders that are brighter than average are widespread, some of which may be of exogenous origin, derived from S-type asteroids \citep{TS20,S21}. The dominant color variations at visible wavelengths are variations in the spectral slope, where some areas are bluer than average, meaning they have a shallower or negative spectral slope \citep{Su19}. Examples of blue terrain are the equatorial ridge, the poles, some impact craters, and a large boulder (Otohime Saxum) at the south pole \citep{M20,TD20,T21,Yo21}. Observations made during the collection of samples suggest that such blue terrain is relatively young \citep{M20}. The near-infrared spectrometer (NIRS3) found the dominant spectral variations in the near-IR to be associated with a narrow absorption feature centered at 2.72~\textmu m, linked to hydroxyl (OH)–bearing minerals \citep{K19}. In addition, weak variations in the spectral slope in the near-IR have been uncovered \citep{G20}. The thermal infrared imager (TIR) found Ryugu's average thermal inertia to be consistent with that of the rock seen by MASCOT \citep{O20,Sh20}. The surface appears to be dominated by highly porous boulders, but the TIR also identified some boulders with a higher density, more similar to that of typical CC meteorites \citep{O20}. Subsequent observations at low altitude revealed the existence of ultra-porous boulders, with a density approaching that of cometary material \citep{Sa21}.

Now that the samples collected by Hayabusa2 have been recovered, we can directly compare the Ryugu particles with the MASCam images. A preliminary analysis of the Ryugu samples indicates that any chondrules and Ca-Al-rich inclusions larger than a millimeter are absent \citep{Ya21}. Based on the color, shape, structure, and near-IR spectral properties of the particles, the samples are considered to be representative of the asteroid surface \citep{T22,P21,Ya21}. As the rock seen by MASCOT features multicolored inclusions, the question of how representative this rock is of Ryugu as a whole becomes important. To answer this question, we put the {in situ} MASCOT measurements in the context of remote observations of the landing site by the Hayabusa2 ONC, NIRS3, and TIR instruments. In the process, we investigated various types of terrains in the landing area that can be recognized in observations by both lander and orbiter.

\section{Data}

\subsection{ONC}

The optical navigation camera ONC-T is a telescopic framing camera with seven narrowband filters (Table~\ref{tab:ONC_filters}) and a field of view of $6.35^\circ \times 6.35^\circ$. The spatial sampling is 22~arcsec per pixel, and the full width at half maximum of the point spread function is less than 2~pixels for all filters \citep{K17}. We analyzed ONC images that are calibrated to level~2d (version~3d), which have the unit of reflectance ($I/F$) \citep{S18,T19,K21}. The observation and illumination geometry in each image pixel, as defined by the incidence, emission, and phase angles, were calculated using Ryugu shape model SHAPE\_SPC\_3M\_v20200323, derived from stereophotoclinometry \citep{Hi20}. We projected the images onto a regular latitude-longitude grid with bins of $0.01^\circ$ (equirectangular projection), using spherical gridding (SPH\_SCAT) in IDL.

We analyzed six multispectral ONC-T image sets acquired around the time of the MASCOT landing (Table~\ref{tab:image_sets}). The first and last set were acquired two months before and six months after the landing, respectively. They contain the highest resolution multicolor images of the landing site available. The first four sets were acquired using all seven filters, whereas for the last two sets the camera used only four filters. The projected images registered well within each set, but there were large differences between the sets. Therefore, we manually shifted the sets in latitude and longitude to match surface features in the landing area. MASCOT itself can be easily identified in the images because of its high reflectance. It is visible in all sets but the first, which was acquired prior to landing. The lander dimensions are $28 \times 29 \times 21$~cm$^3$ \citep{H17}. Its longest side measures 1.0~pixels in images of set~2 (lowest resolution) and 1.5~pixels in images of set~6 (highest resolution).

\begin{table}
        \centering
        \caption{ONC-T filter characteristics \citep{T19}.}
        \label{tab:ONC_filters}
        \begin{tabular}{lll}
                \hline
                \hline
                Filter & $\lambda_{\rm eff}$ & Bandwidth \\
                & (nm) & (nm) \\
                \hline
                \texttt{ul} & 398 & 36 \\
                \texttt{b}  & 480 & 27 \\
                \texttt{v}  & 549 & 31 \\
                \texttt{Na} & 590 & 12 \\
                \texttt{w}  & 700 & 29 \\
                \texttt{x}  & 857 & 42 \\
                \texttt{p}  & 945 & 56 \\
                \hline
        \end{tabular}
        \tablefoot{The effective wavelength was calculated with respect to the solar spectrum.}
\end{table}

\begin{table*}
        \centering
        \caption{ONC-T multiband image set details.}
        \label{tab:image_sets}
        \begin{tabular}{lllllll}
                \hline
                \hline
                Set & Date & Time span (UTC) & Distance (km) & Pixel scale (m) & Phase angle & Filters \\
                \hline
                1 & 1 Aug 2018 & 19:17:57 - 19:19:33 & $5.17 \pm 0.02$ & $0.553 \pm 0.002$ & $18.5^\circ \pm 0.4^\circ$ & \texttt{ul}, \texttt{b}, \texttt{v}, \texttt{Na}, \texttt{w}, \texttt{x}, \texttt{p} \\
                2 & 3 Oct 2018 & 17:15:10 - 17:16:46 & $2.81 \pm 0.01$ & $0.300 \pm 0.001$ & $13.1^\circ \pm 0.3^\circ$ & \texttt{ul}, \texttt{b}, \texttt{v}, \texttt{Na}, \texttt{w}, \texttt{x}, \texttt{p} \\
                3 & 4 Oct 2018 & 00:25:09 - 00:26:45 & $2.79 \pm 0.01$ & $0.298 \pm 0.001$ & $15.7^\circ \pm 0.3^\circ$ & \texttt{ul}, \texttt{b}, \texttt{v}, \texttt{Na}, \texttt{w}, \texttt{x}, \texttt{p} \\
                4 & 4 Oct 2018 & 00:55:09 - 00:56:45 & $2.76 \pm 0.01$ & $0.295 \pm 0.001$ & $12.2^\circ \pm 0.4^\circ$ & \texttt{ul}, \texttt{b}, \texttt{v}, \texttt{Na}, \texttt{w}, \texttt{x}, \texttt{p} \\
                5 & 21 Mar 2019 & 19:05:54 - 19:06:49 & $1.77 \pm 0.01$ & $0.198 \pm 0.001$ & $20.9^\circ \pm 0.4^\circ$ & \texttt{ul}, \texttt{b}, \texttt{v}, \texttt{x} \\
                6 & 25 Apr 2019 & 02:51:14 - 02:52:08 & $1.76 \pm 0.01$ & $0.188 \pm 0.001$ & $27.1^\circ \pm 0.4^\circ$ & \texttt{ul}, \texttt{b}, \texttt{v}, \texttt{x} \\
                \hline
        \end{tabular}
        \tablefoot{Date and time of image acquisition are from the image file name. Distance, pixel scale, and phase angle are for the MASCOT landing site.}
\end{table*}

\subsection{NIRS3}

The near-IR spectrometer NIRS3 is a passively cooled linear image sensor with indium arsenide (InAs) photo diodes, and is sensitive in the 1.8-3.2~\textmu m wavelength range with a field of view of $0.11^\circ$, a spectral sampling of 18~nm per pixel, and a spatial resolution of 40~m at 20~km altitude \citep{I17}. The NIRS3 spectra analyzed in this study are level~2d data, calibrated to reflectance ($I/F$) and corrected for thermal emission. We selected two spectral data sets, one acquired on 19~July 2018 and the other on 30~October 2018 at phase angle $17.7^\circ$ and $7.7^\circ$, respectively. The 19~July set comprises 17~spectra acquired sequentially with mid-exposure times between 10:42:10.6 and 10:42:31.9 UTC. The 30~October set comprises 23~spectra acquired sequentially with mid-exposure times between 09:51:05.6 and 09:51:34.9 UTC. The observation and illumination geometry for each spectral footprint, as defined by the incidence, emission, and phase angles, were calculated using Ryugu shape model SHAPE\_SFM\_200k\_v20200815, derived from the shape-from-motion technique \citep{Hi20}.

\subsection{TIR}

The TIR is a two-dimensional uncooled micro-bolometer array with $328 \times 248$ effective pixels that takes images of thermal IR emission in the 8-12~\textmu m wavelength range. The field of view is $16^\circ \times 12^\circ$ with a spatial resolution of $0.05^\circ$ per pixel, which corresponds to 17~m per pixel at 20~km altitude \citep{O17}. For the thermal analysis we used data calibrated to level~2 (brightness temperature) as archived \citep{Ok21}. The brightness temperature has a typical uncertainty of 3~K \citep{O17}. The shape model used for the thermal simulations is the local digital elevation model (DEM) by \citet{P19}, re-meshed to reduce the number of facets. Prior to analysis, the TIR thermal images were registered to ONC images. The TIR data sets analyzed in this paper were acquired on 21~March 2019 at 19:01:37, 19:05:09, 19:08:41, 19:12:13, 19:29:53, and 19:33:25 UTC (six~sets), as recorded in the file name.

\section{Analysis and results}

\subsection{Terrain classification}

MASCOT was aimed to land in an area just south of the equator around longitude $320^\circ$E that was chosen, among other reasons, for its high potential for spectroscopic diversity \citep{L20}. Of particular interest is ``blue'' terrain, which features a neutral-to-negative spectral slope and is thought to expose fresh material \citep{Su19,M20}. After landing, MASCam made multispectral observations of one rock in particular \citep{J19,S20}, in a location that we refer to as the ``landing site.'' As the descent was uncontrolled and the lander was expected to bounce, the exact location of the landing site could not be predicted. But images acquired by Hayabusa2 at the time of descent were used to determine its coordinates as latitude $22.31^\circ$S and longitude $317.15^\circ$E \citep{P19,S19b,S19a}.

\begin{figure}
        \centering
        \includegraphics[width=.47\textwidth,angle=0]{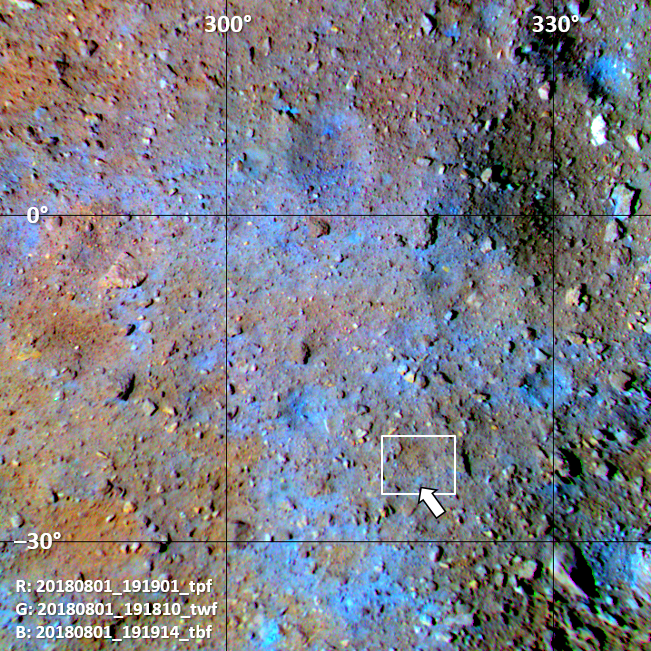}
        \caption{Composite of projected ONC images acquired prior to the MASCOT landing (set~1) with (red, green, blue) = (\texttt{p}, \texttt{w}, \texttt{b}). Colors are saturated. The file names of the images associated with each color channel of the composite are indicated. The white rectangle is the landing area as detailed in Fig.~\ref{fig:evolution}. The arrow points at a small, distinctly blue feature that we identify as an impact crater.}
        \label{fig:context}
\end{figure}

\begin{figure*}
        \centering
        \includegraphics[width=.94\textwidth,angle=0]{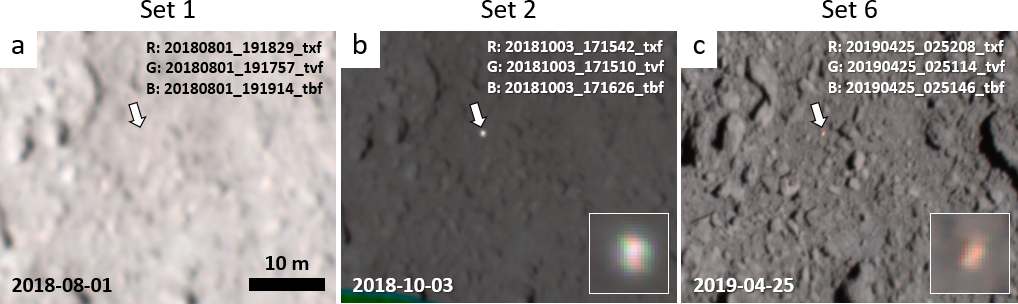}
        \caption{Color composites of identically projected ONC images of the MASCOT landing area. The arrows point at the landing site. ({\bf a})~Before landing (set~1). ({\bf b})~Right after landing, with MASCOT in the nominal orientation (set~2). ({\bf c})~Half a year after landing, with MASCOT in the final orientation (set~6). The file names of the images associated with each color channel of the composite are indicated. The brightness in each composite is scaled from zero to the maximum reflectance over all color channels. The insets show close-ups of MASCOT.}
        \label{fig:evolution}
\end{figure*}

The lander came to rest in what we refer to as the ``landing area'' (indicated in Fig.~\ref{fig:context}). The landing area appears almost devoid of blue terrain, with the only distinctly blue feature being a small spot near the lower boundary, which we will discuss later. Figure~\ref{fig:evolution} zooms in on the landing area, showing color composites of three separate ONC image sets (1, 2, and 6 in Table~\ref{tab:image_sets}). The composites have the ONC \texttt{b} (blue) and \texttt{v} (green) filters in the blue and green color channels, and the \texttt{x} filter in the red channel (Table~\ref{tab:ONC_filters}). Filter \texttt{x} is centered on a wavelength a little beyond red in the visible spectrum, so the color composites are close to natural color, as may be perceived by the eye. Composite (a) (set~1) is made from images acquired two months before the release of MASCOT and shows the landing site at relatively low resolution. The insets in (b) and (c) show MASCOT close up, as it appears in the re-projected (resampled) images. Composite (b) (set~2) is made from images acquired just after landing and has a bright spot at the landing site that corresponds to the lander's highly reflective, white top side \citep{H17,J17}. This was the nominal attitude in which MASCam acquired the images of the rock. The lander size is 1.0~pixels in the original, unprojected images, so neither its shape nor orientation can be distinguished in the inset of (b). Apparent color variations over the face of the lander are artifacts due to projection mismatches in the different spectral bands. Composite (c) (set~6) is made from images acquired half a year after landing. It shows the lander on its side, that being the consequence of its final relocation \citep{S19a}. The inset in Fig.~2c shows a close-up of MASCOT. The lander no longer appears white but rather has assumed an orange color, which derives from the foil that covered its sides \citep{H17,J17}. The size of the lander is 1.5~pixels in the original images, and both its slightly elongated shape and orientation can be recognized, including a faint shadow on the left side. The images in set~5 have a similar pixel scale as those in set~6, and the lander's appearance (color, shape, orientation) in set~5 is the same as that in set~6.

\begin{figure}
        \centering
        \includegraphics[width=.43\textwidth,angle=0]{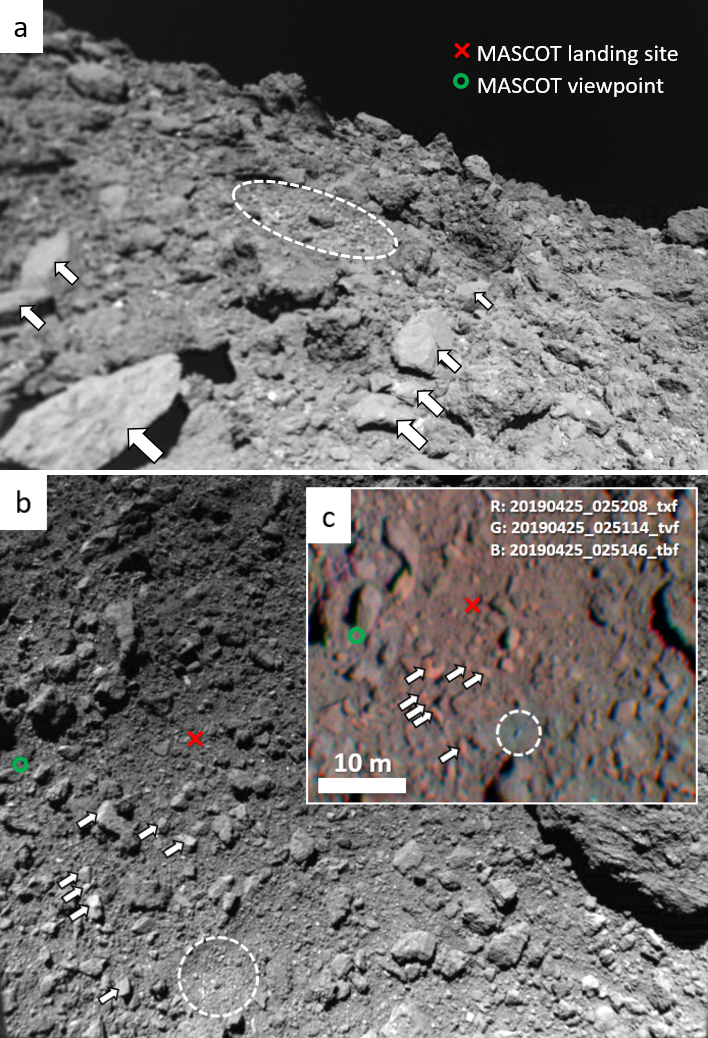}
        \caption{Images of the MASCOT landing area. ({\bf a})~MASCam image \#103, which was the fourth of the descent sequence. ({\bf b})~ONC image 20181003\_015814\_w2f (3 Oct 2018). ({\bf c})~Color composite of projected ONC images from set~6 (Fig.~\ref{fig:evolution}c), with colors strongly saturated and red intensity reduced. The arrows indicate a number of smooth, angular boulders, and the dashed circles outline a 5~m large impact crater. The green circle represents the location of MASCOT at the moment it acquired the image in panel (a) \citep{S19b}.}
        \label{fig:landing_site}
\end{figure}

Unfortunately, MASCam did not image the landing site itself prior to landing. We therefore focus on terrain that was imaged during the descent, located immediately south of the landing site (Fig.~\ref{fig:landing_site}a). Several surface features seen by MASCam can also be identified in an ONC image that was acquired immediately after the release of MASCOT (Fig.~\ref{fig:landing_site}b). We recognize what appears to be a small impact crater of about 5~m in diameter with a boulder at its center (circled). This crater appears blue in Fig.~\ref{fig:landing_site}c and can be identified with the small blue spot at the bottom of the rectangle in Fig.~\ref{fig:context}. The central boulder appears especially blue. Arrows point at a cluster of smooth, bright, and angular boulders that \citet{J19} identified as type~2, in accordance with the definition in \citet{Su19}. These boulders appear distinctly red in the enhanced color composite. Boulders and cobbles with a rough appearance do not stand out in Fig.~\ref{fig:landing_site}c as having a particular color, with most appearing as either average or somewhat reddish or bluish in color. The MASCOT landing site itself (red cross) also appears unremarkable in terms of its color. To summarize, we selected three areas and landforms in the landing area (hereafter called ``units'') for further analysis: (1)~the terrain immediately surrounding the landing site, (2)~a collection of smooth, angular rocks, and (3)~a small impact crater.

\subsection{ONC}

\begin{figure*}
        \centering
        \includegraphics[width=.49\textwidth,angle=0]{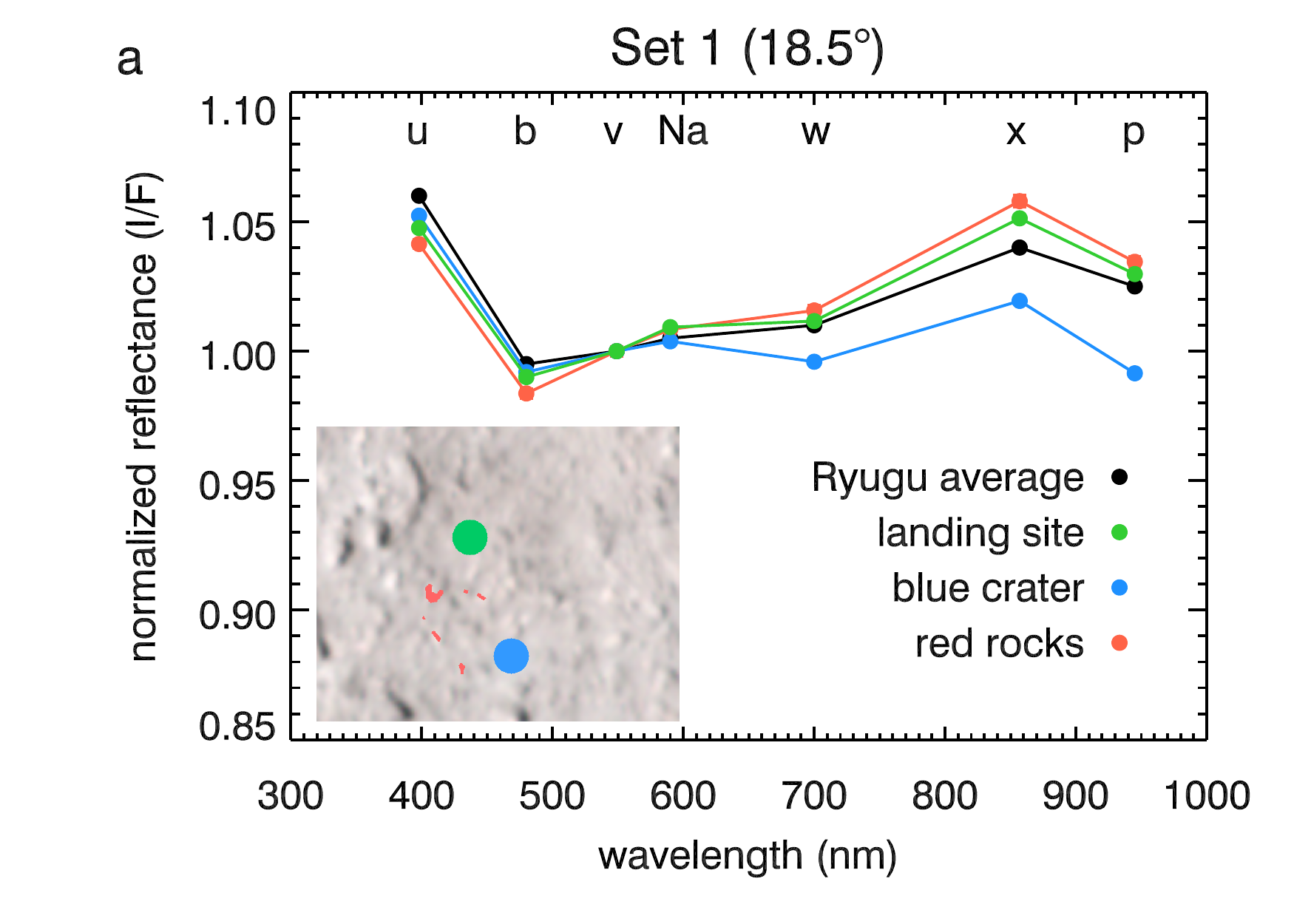}
        \includegraphics[width=.49\textwidth,angle=0]{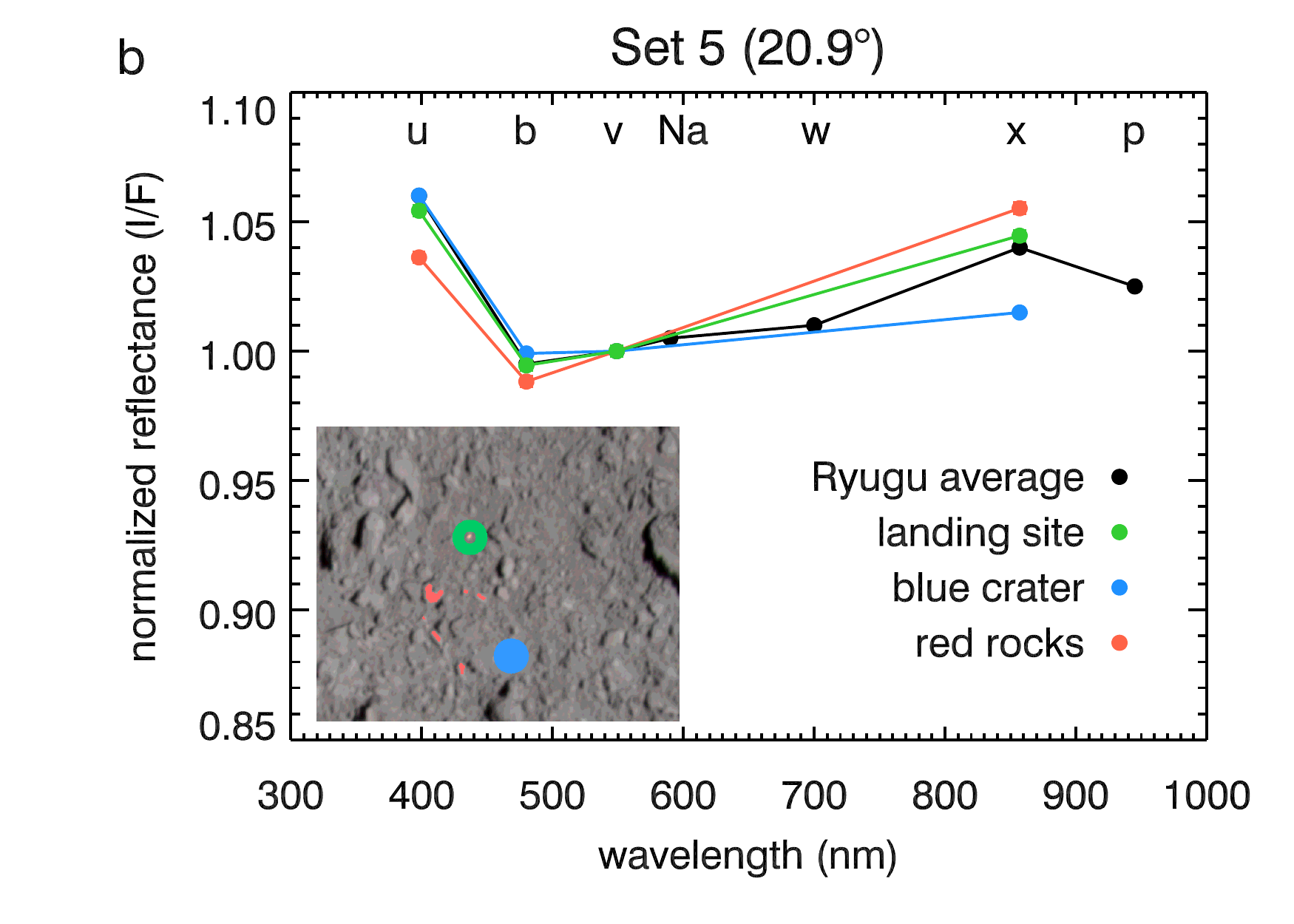}
        \caption{Spectra of different terrain units in the MASCOT landing area for image sets~1 ({\bf a}) and 5 ({\bf b}), with the average phase angle indicated in the plot title. Error bars indicate the standard error of the mean and are generally smaller than the plot symbols. The insets show color composites of the respective sets, including the locations of the pixels that were averaged for each unit in the legend. The ``landing site'' unit for set~5 excludes the lander itself. Spectra are normalized at 550~nm, and letters correspond to ONC filter names. The Ryugu average spectrum is that for the $19.5^\circ$ phase angle in Fig.~4 of \citet{TD20}.}
        \label{fig:spectral}
\end{figure*}

We characterized the spectral properties of the three units by extracting spectra from the ONC image sets listed in Table~\ref{tab:image_sets}. As the absolute reflectance is generally a function of phase angle, we compared image sets that were acquired at approximately the same phase angle. Sets~1 and 5 were both acquired around phase angle $20^\circ$, with a difference of only $2.4^\circ$ between the sets. Figure~\ref{fig:spectral} shows the reflectance spectra for each unit as the average over the areas as defined in the insets. The resolution of the shape model we used for projecting the images is not sufficient to allow for the complete removal of photometric effects due to topography by means of photometric correction. Because the absolute reflectance of the averaged units may be affected by differences in topography, we normalized the spectra at 550~nm to better assess the spectral shape. Because the reflectance also varies within each unit due to topography, we adopted the standard error of the mean instead of the standard deviation for the error bars for the data points; we mostly assessed the trustworthiness of the spectral shapes from similarities between the two sets. Set~1 contains all eight spectral bands that are available to the ONC, whereas set~5 lacks three of them. For reference, Fig.~\ref{fig:spectral} also includes the average spectrum of the Ryugu surface at phase angle $19.5^\circ$ (labeled ``Ryugu average'') derived by \citet{TD20}. The spectral differences between the units are small but generally consistent in both sets. The reflectance spectrum of the landing site (unit~1) is close to that of average Ryugu, being slightly lower at the blue end of the spectrum and slightly higher at near-IR wavelengths. The smooth rocks (unit~2) are redder than average Ryugu (steeper spectral slope) over the full wavelength range in both sets. The impact crater (unit~3) is bluer than average Ryugu, with a reflectance that is markedly lower at near-IR wavelengths in both sets. Although the phase angles of the other four image sets are not as close to the $19.5^\circ$ of the average Ryugu spectrum, their spectral plots all confirm the color trends (Fig.~\ref{fig:appendix}).

\subsection{NIRS3}

Our analysis of NIRS3 spectral data is focused on three spectroscopic quantities. For the definition of these quantities we considered the reflectance $r_\lambda$ ($I/F$) at central wavelength $\lambda$ in \textmu meters. The three reflectance values, $r_{1.9}$, $r_{2.0}$, and $r_{2.5}$, are defined as the average reflectance of three NIRS3 channels, one at the central wavelength and its two neighbors. The associated uncertainties, $\sigma_{1.9}$, $\sigma_{2.0}$, and $\sigma_{2.5}$, are defined as the standard deviation of the three channels. The reflectance at the center of the 2.72~\textmu m absorption band is defined in a different way because the band is so narrow. We estimate $r_{2.72}$ as the reflectance of the NIRS3 channel at the central wavelength, and uncertainty $\sigma_{2.72}$ as the average of $\sigma_{1.9}$, $\sigma_{2.0}$, and $\sigma_{2.5}$. The three spectroscopic quantities that we analyzed are (1)~the reflectance at 2.0~\textmu m, defined as $r_{2.0}$, (2)~the near-IR spectral slope, $s$, with uncertainty $\sigma_s$ (both in \textmu m$^{-1}$), defined as \citep{G20}
\begin{equation}\label{eq:spectral_slope_mean}
s = \frac{r_{2.5} - r_{1.9}}{(2.5 - 1.9) r_{1.9}},
\end{equation}
with
\begin{equation}\label{eq:spectral_slope_stdev}
\sigma_s = \frac{(\sigma_{1.9}^2 + \sigma_{2.5}^2)^{1/2}}{(2.5 - 1.9) r_{1.9}},
\end{equation}
and (3)~the depth of the 2.72~\textmu m absorption band, $d$, with uncertainty $\sigma_d$, defined as\begin{equation}\label{eq:band_depth_mean}
d = \frac{r_{2.5} - r_{2.72}}{r_{2.5}},
\end{equation}
with
\begin{equation}\label{eq:band_depth_stdev}
\sigma_d = \frac{(\sigma_{2.5}^2 + \sigma_{2.72}^2)^{1/2}}{r_{2.5}}.
\end{equation}
As $s$ and $d$ are compound quantities, $\sigma_s$ and $\sigma_d$ are large compared to $\sigma_{2.0}$.

\begin{figure*}
        \centering
        \includegraphics[width=.43\textwidth,angle=0]{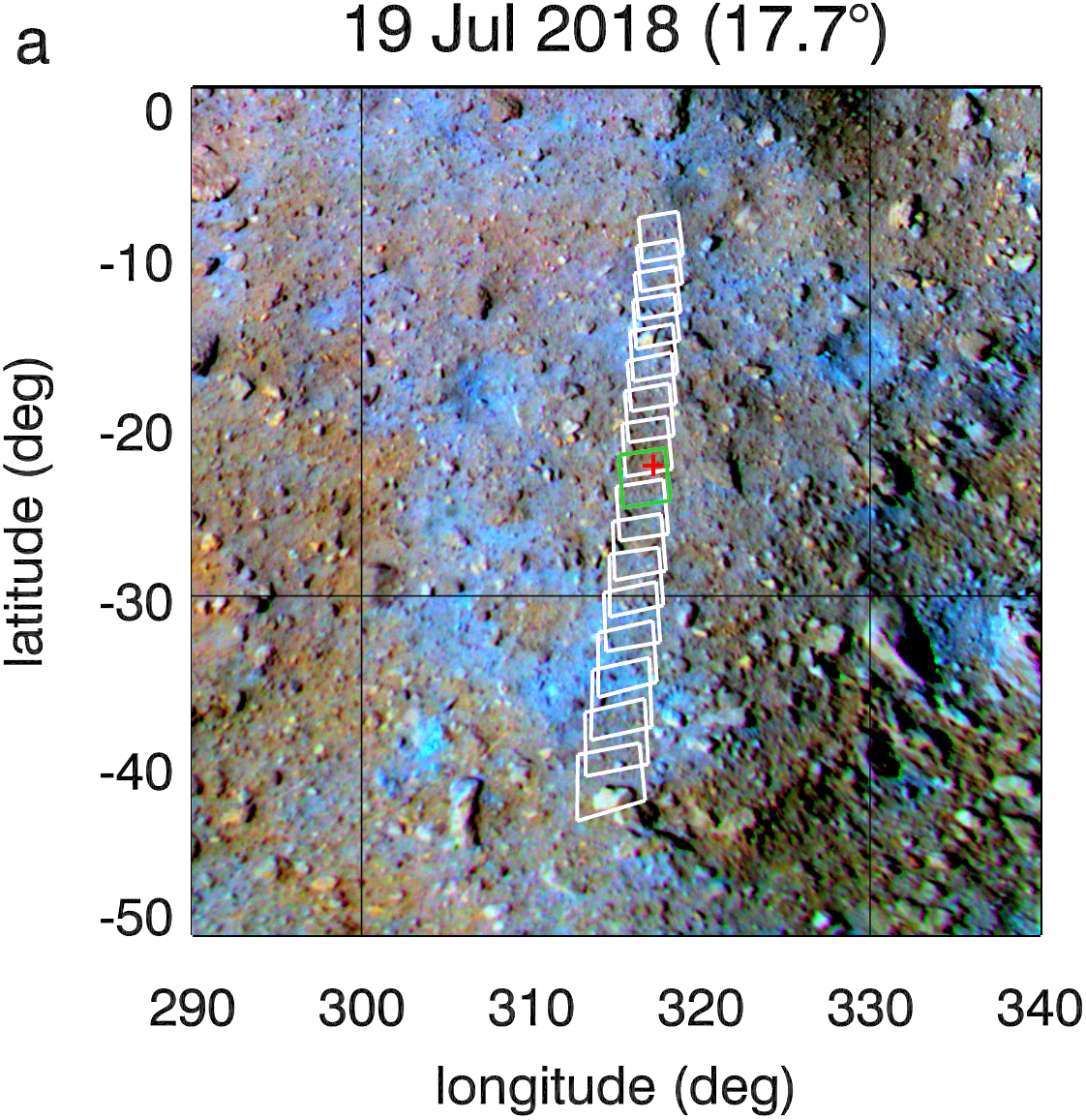}
        \includegraphics[width=.43\textwidth,angle=0]{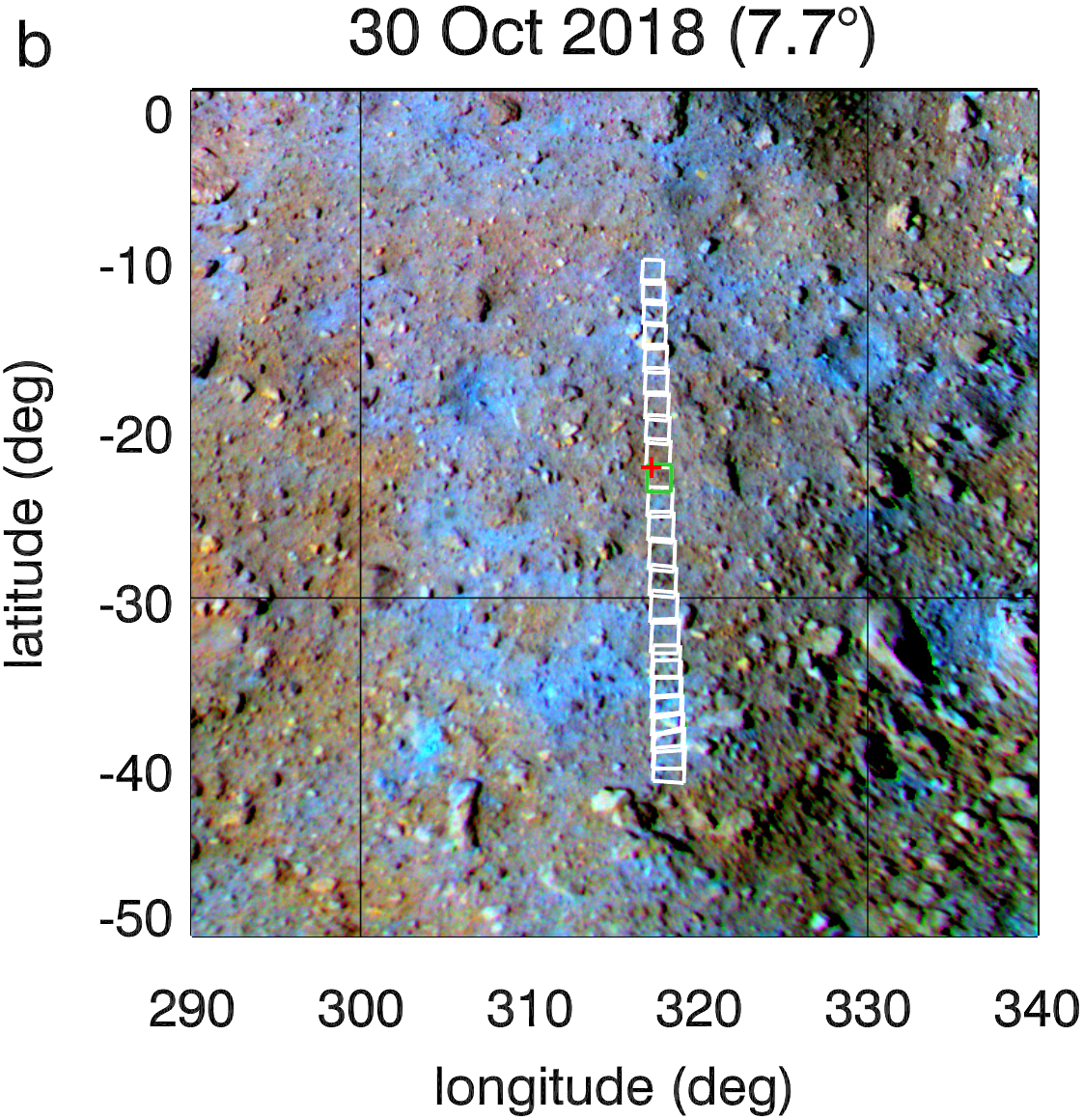}
        \caption{Footprints of two NIRS3 transects that run across the MASCOT landing area (\textcolor{red}{+}), overlaid on the enhanced ONC color composite in Fig.~\ref{fig:context}. ({\bf a})~Spectra acquired at phase angle 17.7° on 19 July 2018. ({\bf b})~Spectra acquired at phase angle 7.7° on 30 October 2018. Footprints that cover the landing site are marked in green.}
        \label{fig:NIRS3_areas}
\end{figure*}

\begin{figure*}
        \centering
        \includegraphics[width=.46\textwidth,angle=0]{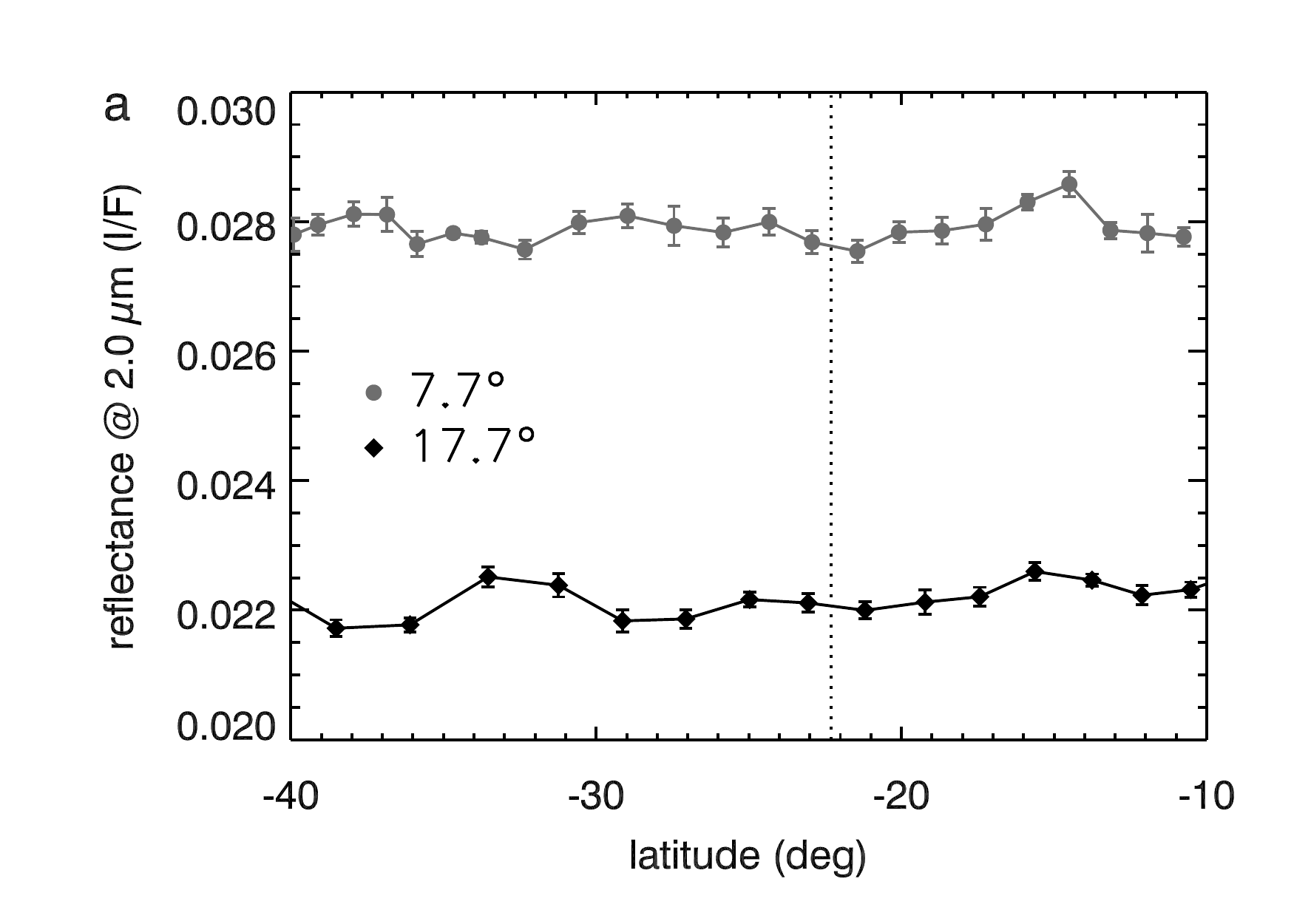}
        \includegraphics[width=.46\textwidth,angle=0]{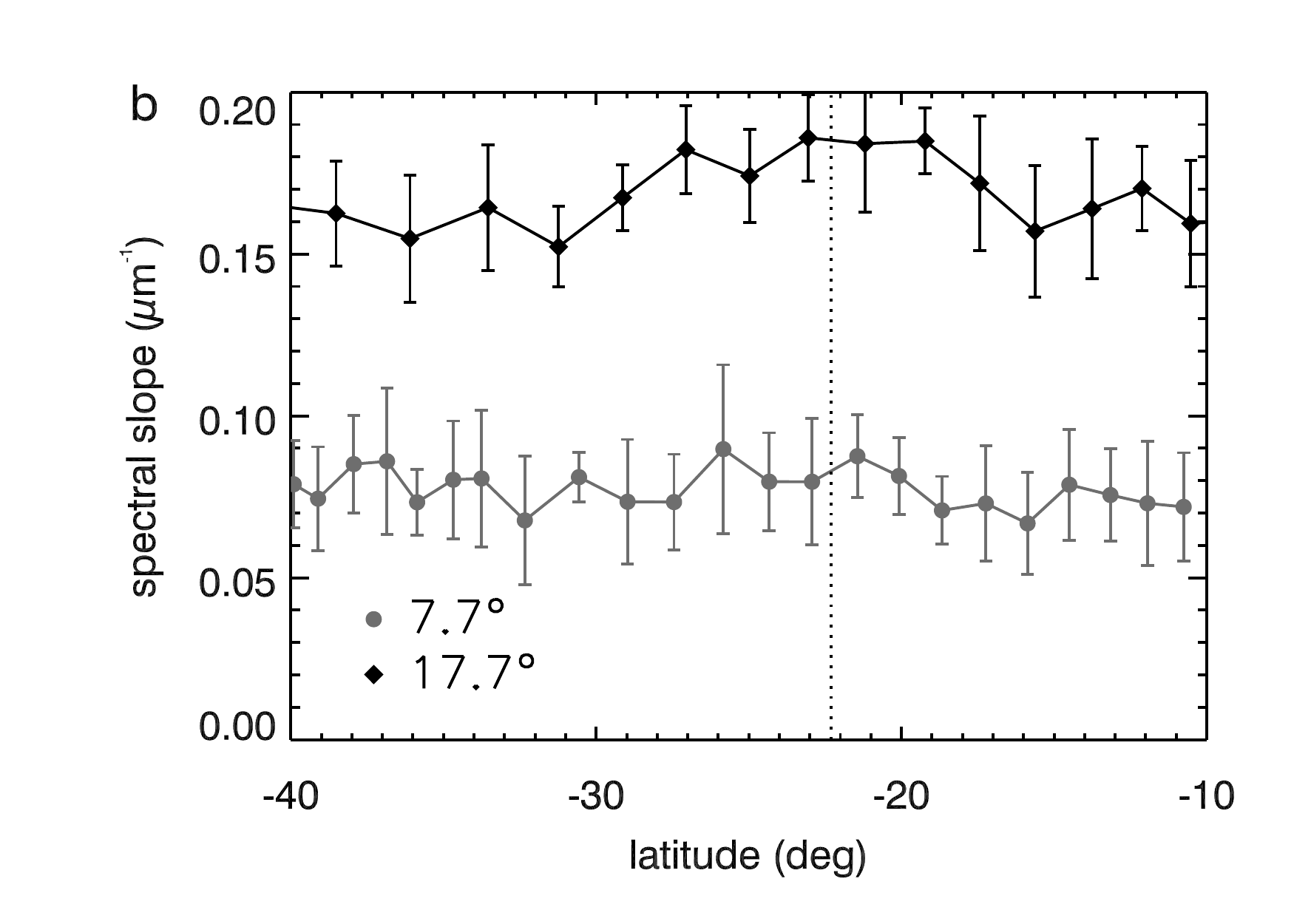}
        \includegraphics[width=.46\textwidth,angle=0]{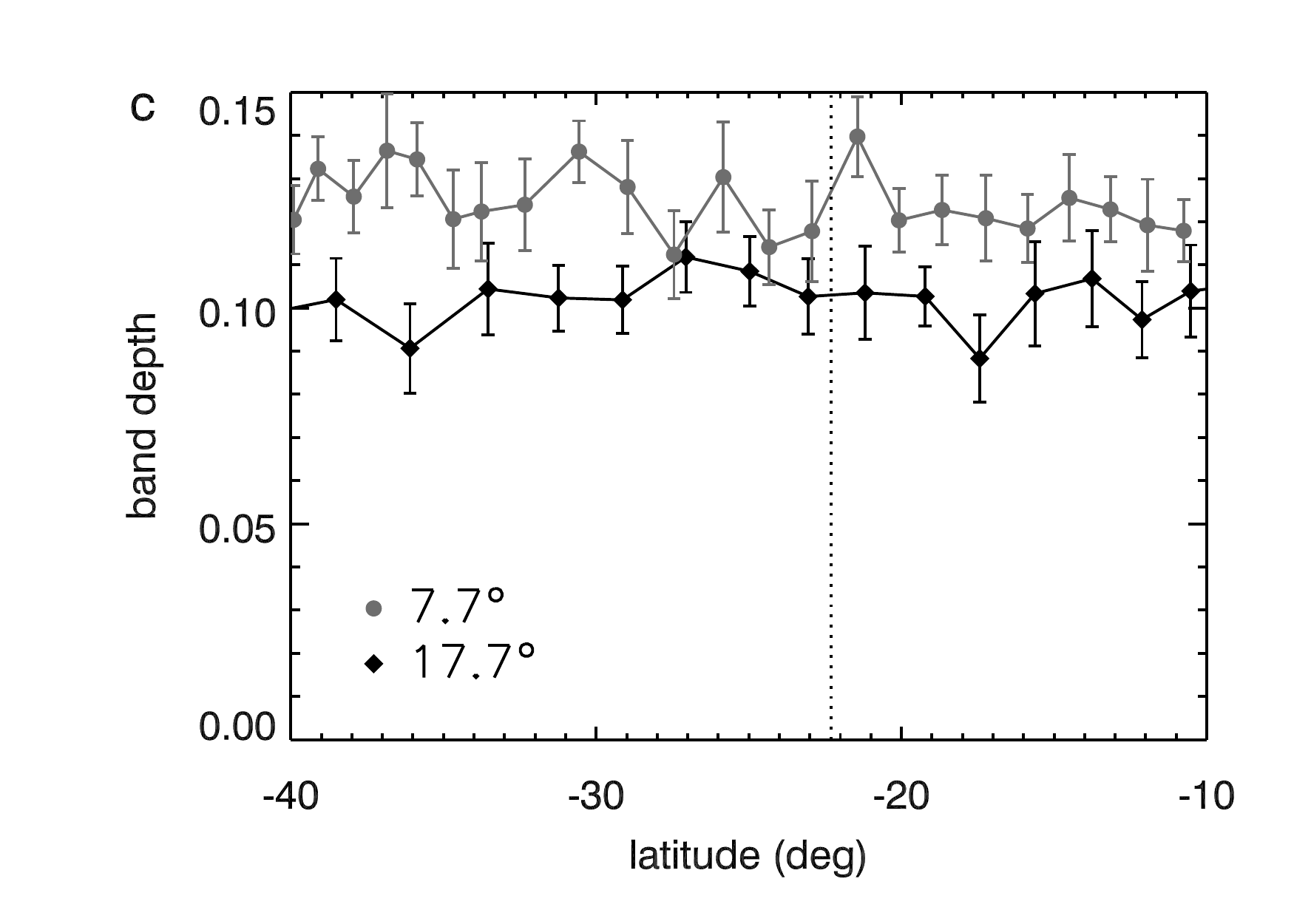}
        \includegraphics[width=.46\textwidth,angle=0]{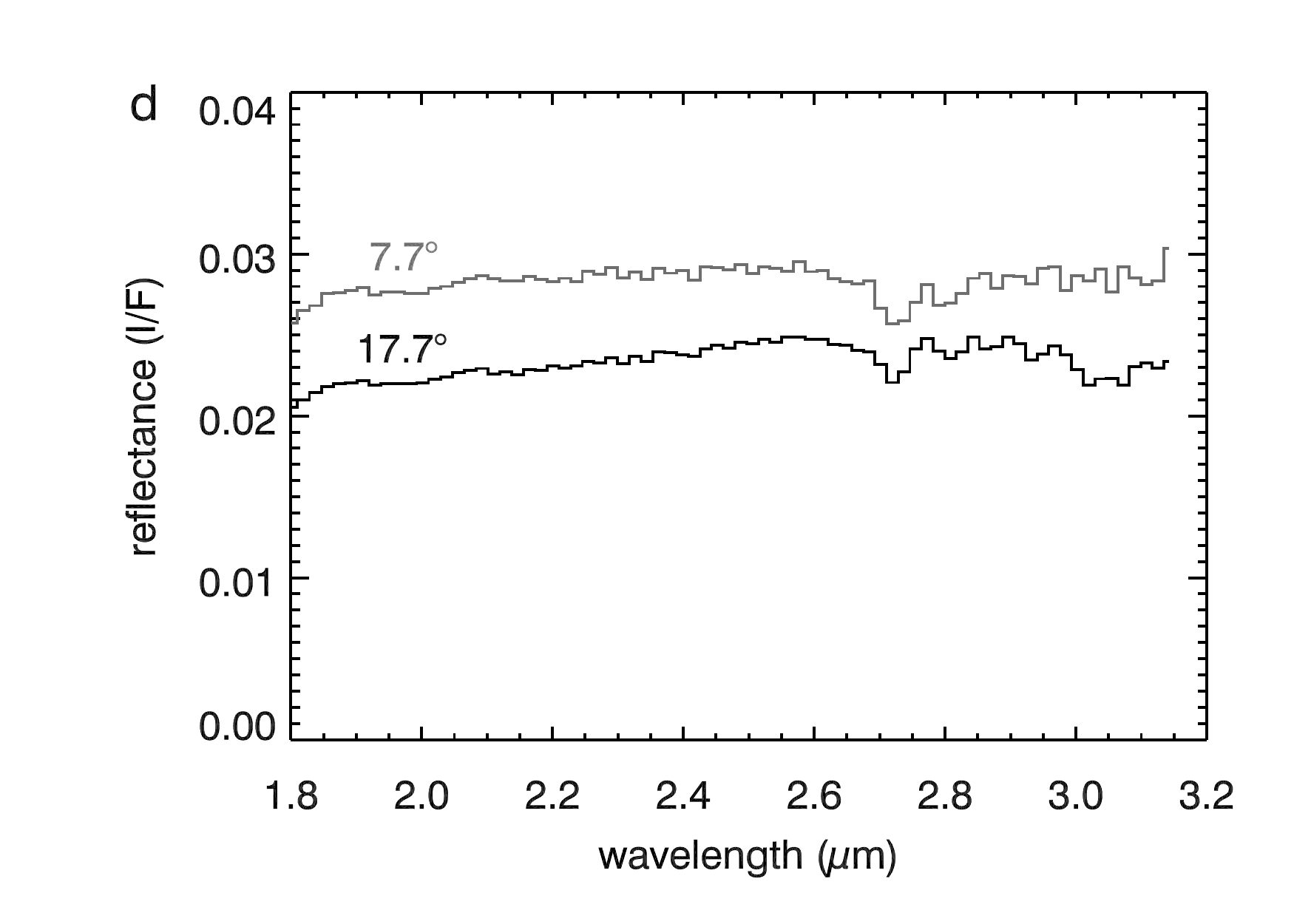}
        \caption{Spectral properties of the MASCOT landing area from the NIRS3 transects in Fig.~\ref{fig:NIRS3_areas}. ({\bf a})~Reflectance at 2.0~\textmu m along the transects. ({\bf b})~Near-IR spectral slope along the transects. ({\bf c})~Depth of the 2.72~\textmu m band along the transects. ({\bf d})~Spectra associated with footprints that cover the landing site, marked as green in Fig.~\ref{fig:NIRS3_areas}. The dotted line in panels (a)-(c) indicates the latitude of the landing site.}
        \label{fig:NIRS3_transects}
\end{figure*}

We identified two NIRS3 transects of high spatial resolution that traverse the MASCOT landing area. Their footprints are shown in Fig.~\ref{fig:NIRS3_areas} on a background that is the ONC enhanced color composite shown earlier. Apart from the landing site itself, at latitude $-22.3^\circ$, the transects also cover relatively blue terrains around latitudes $-15^\circ$ and $-35^\circ$. The two major differences between the transects are the phase angle and spatial resolution. The average phase angle of the transects acquired on 19~July and 30~October 2018 is $17.6527^\circ \pm 0.0001^\circ$ and $7.6831^\circ \pm 0.0002^\circ$, respectively. The spatial resolution of the $17.7^\circ$ transect is about half that of the $7.7^\circ$ transect. The transects run approximately south to north in a straight line. The average longitude of the footprint center of the $17.7^\circ$ and $7.7^\circ$ transects is $316.5^\circ \pm 0.9^\circ$ and $318.0^\circ \pm 0.3^\circ$, respectively. While the upper parts of the transects mostly cover the same terrain, they diverge at the bottom. Nevertheless, to facilitate our analysis we plot the three spectroscopic quantities as a function of only latitude in Fig.~\ref{fig:NIRS3_transects}. Major differences between the transects are due to the difference in average phase angle ($7.7^\circ$ versus $17.7^\circ$). Spectra in the transect with the higher average phase angle have a lower average 2.0~\textmu m reflectance, a higher near-IR spectral slope (redder), and a slightly lower 2.72~\textmu m band depth. The difference in reflectance is consistent with the general shape of the phase curve (stronger shadows at higher phase angle). The increase in spectral slope with phase angle is known as phase reddening and is consistent with both ONC and NIRS3 observations \citep{TD20,D21}. The 2.72~\textmu m band appears to be shallower at larger phase angles, which had not yet been reported for Ryugu. This behavior is different from that observed for the dwarf planet Ceres, also a body of low albedo, for which several bands deepen with phase angle \citep{Lo19}. Across the transects, the 2.0~\textmu m reflectance shows variations that are significant because of the relatively small error bars. In particular, the blue terrains north and south of the landing site stand out as slightly more reflective than average. In contrast, variations in the near-IR slope and the 2.72~\textmu m band depth are on the order of the size of the error bars. This means that the bright, blue terrains do not show a significantly different slope or band depth.

The latitude of the MASCOT landing site is indicated in Fig.~\ref{fig:NIRS3_transects}. The spectral properties of the landing site do not stand out as unusual in the transect plots. Figure~\ref{fig:NIRS3_transects}d shows the spectra associated with the two green footprints in Fig.~\ref{fig:NIRS3_areas}, which represent the area surrounding the landing site. Differences between the two spectra are primarily due to the different phase angle of acquisition, in particular the increase in the spectral slope below 2.6~\textmu m (known as phase reddening) and the decrease in the 2.7~\textmu m band depth.

\subsection{TIR}

\begin{figure}
        \centering
        \includegraphics[width=.43\textwidth,angle=0]{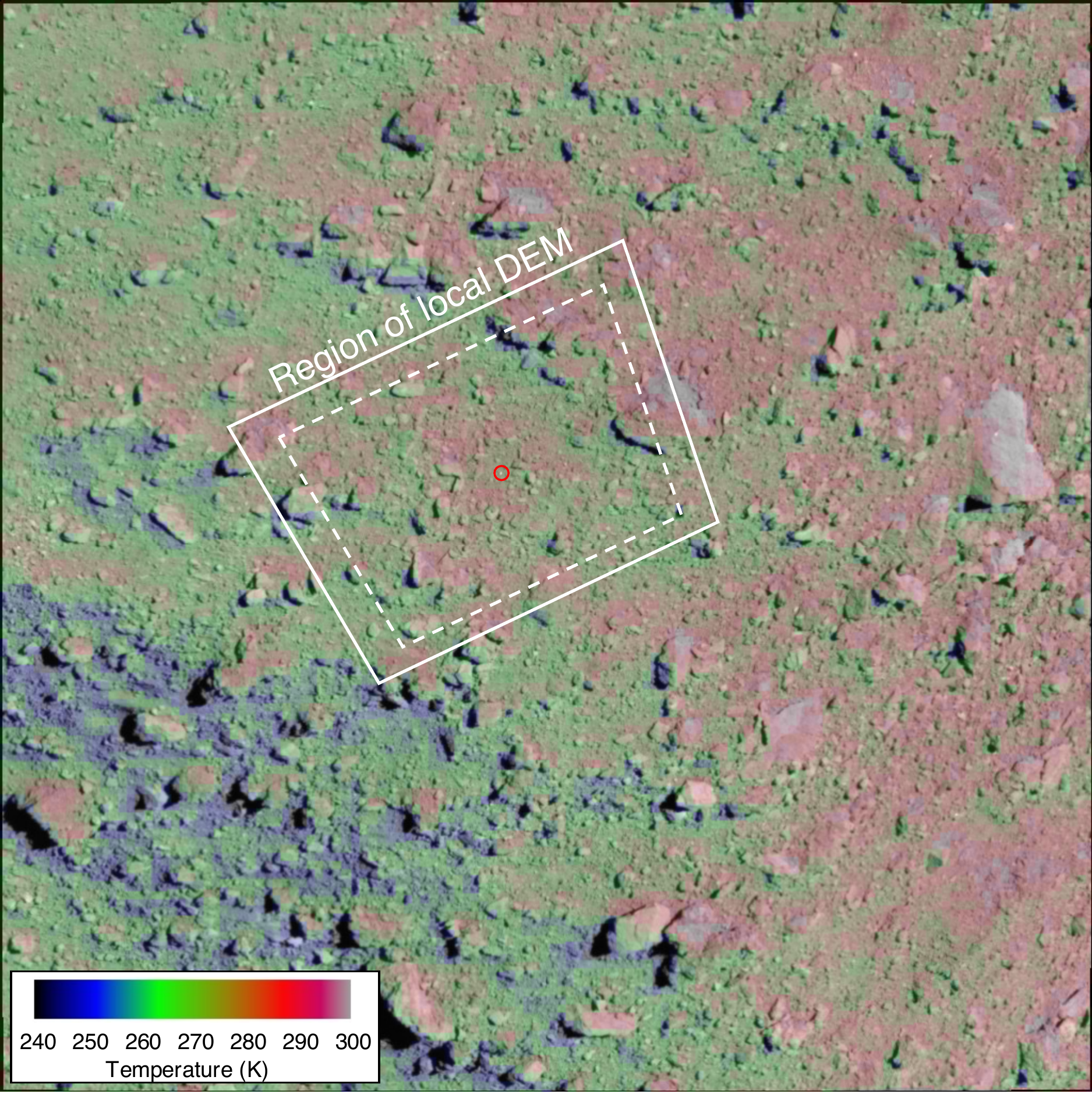}
        \caption{TIR brightness temperature image 20190321\_190509\_l2 registered to ONC-T image 20190321\_190554\_tvf. The bright spot inside the red circle is the MASCOT lander. The solid white box marks the region covered by the local DEM for which we performed the thermal simulations. The dashed white box marks the region for which we derived the thermal inertia (see Fig.~\ref{fig:TIR_inertia}).}
        \label{fig:TIR_temp}
\end{figure}

\begin{figure*}
        \centering
        \includegraphics[width=\textwidth,angle=0]{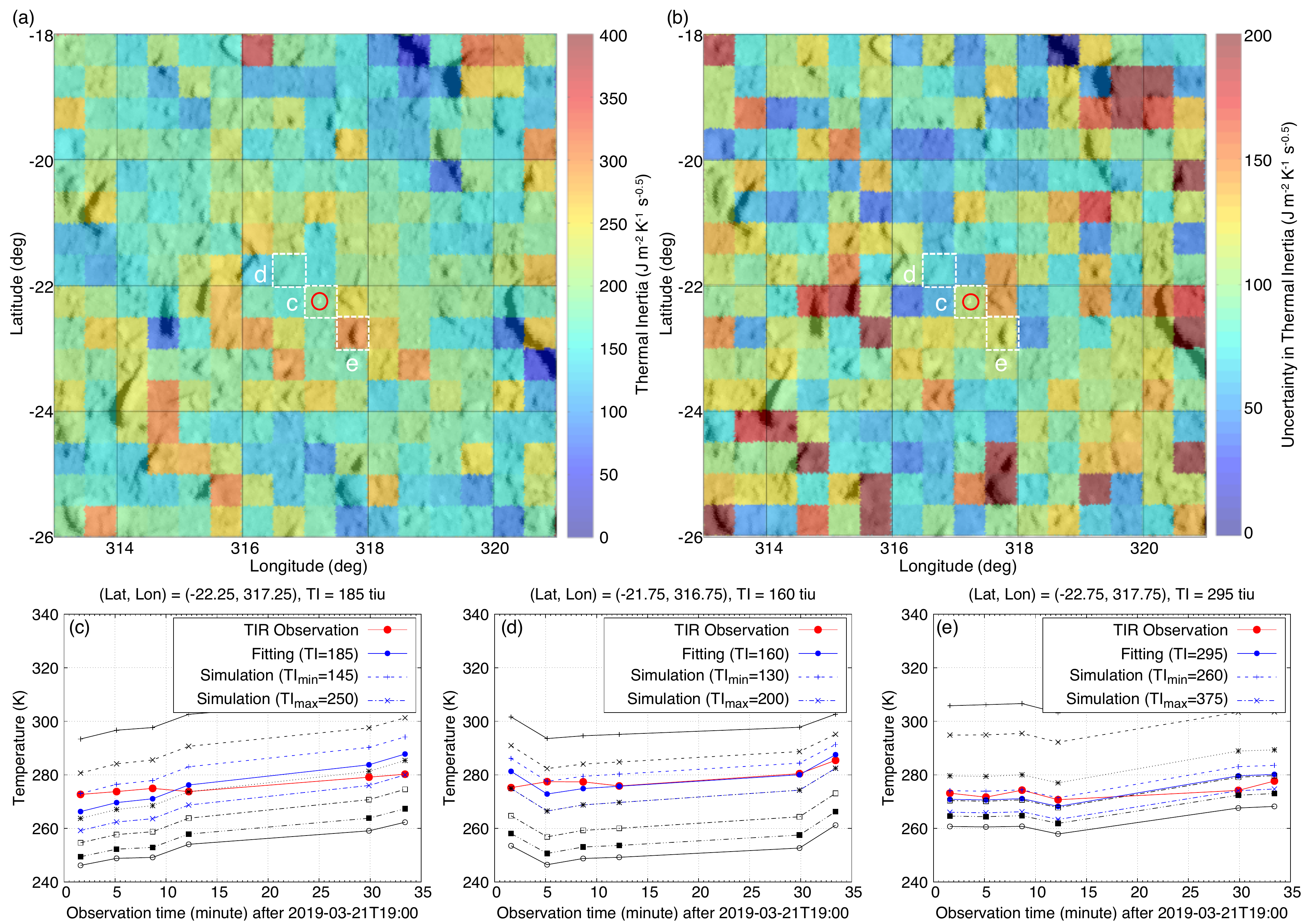}
        \caption{Thermal inertia of the MASCOT landing region. ({\bf a})~Thermal inertia map. The background image is the same ONC-T image as in Fig.~\ref{fig:TIR_temp}. The MASCOT landing site is indicated by a red circle. ({\bf b})~Map of the thermal inertia uncertainty, defined as the difference between the maximum and minimum possible thermal inertia values. ({\bf c-e})~Temperature history of the three grid cells highlighted with a dashed line in panels~{\bf a} and {\bf b}. The red curves are the observations, and others are simulations. In each plot, the black curves represent the simulated temperature with thermal inertia values of 50, 100, 200, 300, 400, 500, and 750~J m$^2$ K$^{-1}$ s$^{-0.5}$. The solid blue curve represents the (interpolated) simulated temperature profile with the best-guess thermal inertia, for which the RMS in Eq. \ref{eq:tir_rms} is minimum, whereas the two dashed blue curves are the profiles with the minimum and maximum possible thermal inertia.}
        \label{fig:TIR_inertia}
\end{figure*}

To analyze the TIR data we used a thermal model developed by \citet{T17}, which was used previously by \citet{O20} and \citet{Sh20}. The analysis comprised the following steps: (1)~We performed thermal simulations using the local DEM with constant thermal inertia ($\Gamma$) values of 50, 100, 200, 300, 400, 500, and 750~J m$^2$ K$^{-1}$ s$^{-0.5}$. The simulations were restricted to the latitude $(-27^\circ, -17^\circ)$ and longitude ($312^\circ, 322^\circ)$ intervals (Fig.~\ref{fig:TIR_temp}, solid white box). (2)~For each observation epoch, we created seven~simulated TIR images based on the results of the thermal simulations and the estimated spacecraft position and attitude \citep{Sa21}. From these seven~simulated images, we created 134 (linearly) interpolated images for a total of 141~images for $\Gamma_k = (50, 55, \ldots, 750)$~J m$^2$ K$^{-1}$ s$^{-0.5}$. (3)~Because of the uncertainty in the position and attitude of the spacecraft, the simulated images ($T_{\rm sim}$) are not perfectly aligned with the observed images ($T_{\rm obs}$). Thus, we performed image alignment by affine transformation of the simulated images, which also provides the latitude and longitude for each pixel of the TIR images \citep{K21b}. (4)~The observed and simulated images were binned into a grid with a 0.5$^\circ$ cell size, corresponding to roughly two~TIR image pixels. We restricted the following analysis to the latitude $(-26^\circ, -18^\circ)$ and longitude ($313^\circ, 321^\circ)$ intervals (Fig.~\ref{fig:TIR_temp}, dashed white box) because the simulated temperatures in the outer regions are affected by thermal input from terrains beyond the DEM. For each grid cell $(i,j)$, we calculated the root-mean-square (RMS) of the residuals of the surface temperature as
\begin{equation}
        \mathrm{RMS}(i,j,\Gamma_k) = \sqrt{\frac{1}{N(i,j)} \sum_{n}^{N(i,j)} [T_{\rm obs}(n,i,j) - T_{\rm sim}(n,i,j,\Gamma_k)]^2},
        \label{eq:tir_rms}
\end{equation}
with $N(i,j)$ the total number of observations in the cell. The $\Gamma_k$ value associated with the lowest RMS was adopted as the thermal inertia $\Gamma(i,j)$ of cell $(i,j)$. In addition, we determined the best-matching thermal inertia for each individual observation by evaluating $T_{\rm obs}(n,i,j) - T_{\rm sim}(n,i,j,\Gamma_k)$ for each $\Gamma_k$, and determined the highest and lowest thermal inertia values of the $N$ observations. We define the uncertainty interval of $\Gamma(i,j)$ as the range spanned by these highest and lowest values. The calculation of the RMS in Eq.~\ref{eq:tir_rms} does not account for the 3~K uncertainty of $T_{\rm obs}$. As this uncertainty has a random nature, it will hardly affect the minimum of the RMS, through which we find $\Gamma(i,j)$. Also, the uncertainty of $T_{\rm obs}$ is small compared to the mismatches between $T_{\rm obs}(n)$ and $T_{\rm sim}(n,\Gamma)$ for the best-matching $\Gamma$ (Fig.~\ref{fig:TIR_inertia}c-e, compare red and blue drawn lines), from which we derive the boundaries of the $\Gamma(i,j)$ uncertainty interval. While the uncertainty of $T_{\rm obs}$ adds uncertainty to the boundaries of the $\Gamma(i,j)$ uncertainty interval, our method to estimate the boundaries themselves remains reasonable.

Figure~\ref{fig:TIR_inertia}a shows a map of the thermal inertia of the MASCOT landing site area, with a map of the associated uncertainty in Fig.~\ref{fig:TIR_inertia}b. The TIR data that cover the landing site itself are most consistent with a thermal inertia of $180^{+65}_{-40}$ J m$^2$ K$^{-1}$ s$^{-0.5}$ (Fig.~\ref{fig:TIR_inertia}c). This value is consistent with the landing site thermal inertia estimate of $200 \pm 7$~J m$^2$ K$^{-1}$ s$^{-0.5}$ from global Ryugu observations by the TIR \citep{Sh20}, but somewhat lower than the $256^{+4}_{-3}$~J m$^2$ K$^{-1}$ s$^{-0.5}$ found by MARA \citep{H22}, possibly because MARA observed only a single boulder \citep{Sh20}. Furthermore, the average thermal inertia of the entire region of Fig.~\ref{fig:TIR_inertia}a is 181~J m$^2$ K$^{-1}$ s$^{-0.5}$ with an average uncertainty of 102~J m$^2$ K$^{-1}$ s$^{-0.5}$, which is consistent with the Ryugu global average of $225 \pm 45$~J m$^2$ K$^{-1}$ s$^{-0.5}$ \citep{Sh20}. Thus, we conclude that the thermophysical properties of the MASCOT landing site are typical for Ryugu.

\section{Discussion}

In this paper we addressed the question of whether the rock in front of MASCOT can be considered representative of Ryugu by analyzing Hayabusa2 observations of the landing area. While the rock itself was too small to be characterized, the landing site has spectral and thermophysical properties that are very close to those of average Ryugu. Two other landforms that we studied are a collection of smooth, angular rocks that appear relatively bright in MASCam images and a small impact crater. The smooth rocks also appear bright in ONC images and are redder than average over the entire visible to near-IR wavelength range. \citet{Su19} originally defined bright and smooth rocks as ``type~2.'' The authors compared the spectral properties of type-2 and type-1 boulders, the latter being dark and rugged. Type-2 boulders were found to be either of similar color or bluer than type-1 boulders. This result is not consistent with the reddish color of the smooth boulders in the landing area. Differences in surface roughness can lead to differences in brightness, but not color, for rocks of the same material. At the special observation geometry known as opposition, when shadows are absent, differences in the intrinsic reflectivity of materials (normal albedo) are readily apparent. \citet{Yo21} studied variations in normal albedo over the surface, but did not address the reflective properties of type-2 rocks. The low surface roughness may explain why type-2 rocks appear relatively bright to MASCam, but they are not necessarily made of intrinsically bright material. \citet{TS20} studied the spectral properties of bright boulders, defined as having a peak normal albedo value larger than 1.5 times Ryugu’s globally averaged normal albedo. A few boulders in their sample have a relatively red spectrum, similar to that of our unit~2 boulders. The authors uncovered the existence of bright boulders with spectra that are similar to those of S-type asteroids. The unit~2 spectrum is inconsistent with that of S-type boulders, as the reflectance does not drop beyond 0.7~\textmu m.

The small crater in the landing area (unit~3) has a relatively fine-grained surface with a small boulder at the center, and was spotted by MASCam just south of the landing site. Its morphology confirms that this roughly circular feature is an impact crater, consistent with a blue color in ONC images. Small, blue impact craters such as these are thought to have formed relatively recently in Ryugu's history \citep{M20}.

\begin{figure}
        \centering
        \includegraphics[width=.47\textwidth,angle=0]{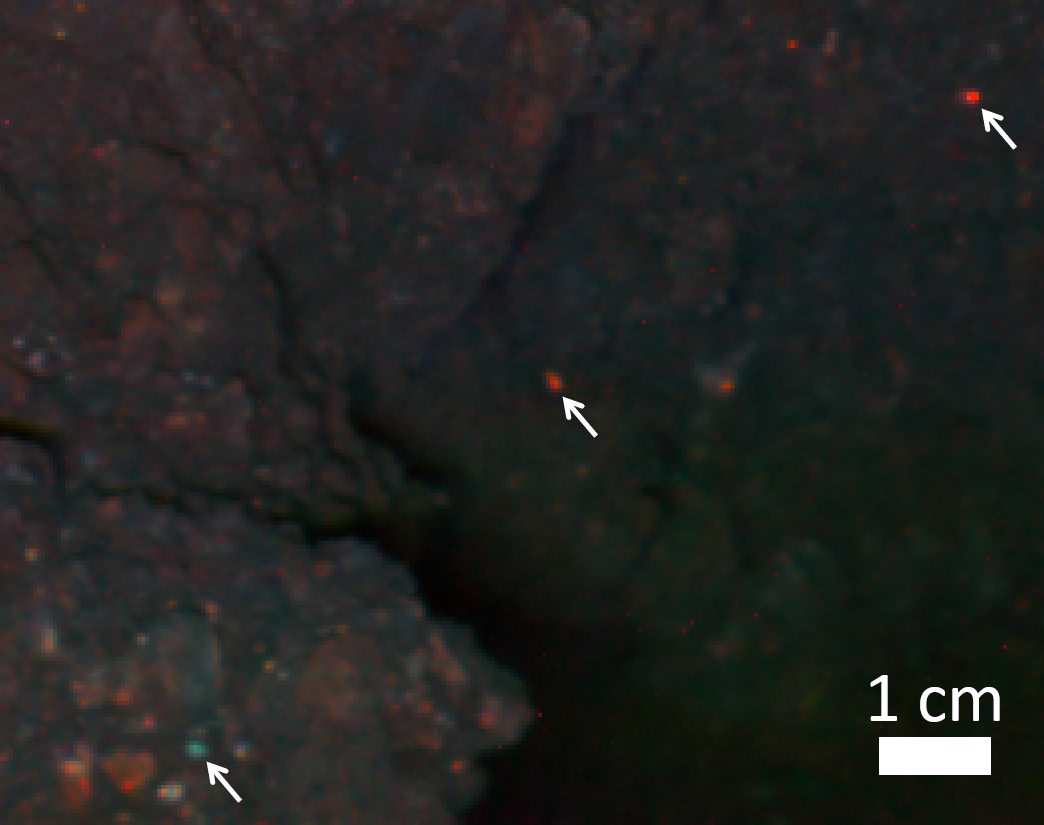}
        \caption{Inclusions in a rock on Ryugu as seen in a MASCam color composite with (red, green, blue) = $({\mathbf I}, {\mathbf G}, {\mathbf B})$ \citep{S20}. The arrows point at a few brightly colored examples. Colors are saturated, and brightness is scaled linearly from zero to maximum (no pixels were overexposed in this cropped image).}
        \label{fig:inclusions}
\end{figure}

Most lines of evidence suggest that the MASCOT rock is indeed representative of Ryugu's surface. But the rock features multicolored inclusions \citep{J19,S20}, whereas the sample particles, apparently, do not. Ryugu particles harbor neither chondrules nor Ca-Al-rich inclusions larger than about a millimeter, although their microporosity is consistent with the estimate for the rock \citep{Ya21}. The samples have unusual reflective properties \citep{P21,Y22}. They show significantly high inter-particle and intra-particle reflectance variations due to the presence of very bright spots that are smaller than a millimeter \citep{Y22}. The spots have a reddish reflectance ten times higher than average, apparently due to specular reflection. The authors speculated that most of the red ``inclusions'' in the MASCOT rock (but none of the blue) can be identified with such specular reflections. While this mechanism may indeed lead to illusionary inclusions, MASCam achieved submillimeter resolution and none of the suspected resolved inclusions appear to be ten times brighter than the rock matrix \citep{S20}. The influence of surface roughness was investigated by \citet{O21b}, who used a spare MASCam to image CC meteorites with both fractured (rough) and cut (smooth) surfaces in a similar illumination geometry as on Ryugu. The authors demonstrated that small-scale topography can lead to brightness enhancements that masquerade as inclusions. Specular reflection was observed, but only rarely. On the Ryugu rock, there were only a handful of the brightest and most strongly colored spots on a total area of roughly $13 \times 13$~cm$^2$ \citep{S20}, with some examples shown in Fig.~\ref{fig:inclusions}. If they were due to specular reflections on a rough surface, we would expect them to be ubiquitous.

The apparent discrepancy between the MASCOT rock and the Ryugu samples may simply illustrate the diversity of rocks present on the surface. After all, rocks with a variety of spectral and thermal properties have been identified, possibly originating from different parent bodies or different layers in a single parent body \citep{TS20,Sa21}. Alternatively, the discrepancy may derive from the sampling method, which involved shooting a pellet into the Ryugu surface \citep{M20}. The samples display compositional heterogeneities at submillimeter scale. Among the bulk of dark particles, \citet{P21} identified several bright particles of about 0.5~mm in size that harbor carbonates or NH-rich compounds. Some carbonate-rich grains feature a steep red slope at wavelengths <1.6 µm, possibly due to absorption by Fe$^{2+}$. These characteristics match those of the aforementioned brightest and most strongly colored spots on the MASCOT rock, although it is unknown what color these bright particles of unusual composition have in the visible wavelength range. Such bright particles may be embedded in the rocks and pebbles of Ryugu, and may end up as individual particles in the samples after being dislodged upon the pellet impact or during atmospheric entry. Inclusions may have originally formed within rocks as a result of aqueous alteration of the parent body or accumulated in rocks due to brecciation. We note that breccias appear to be common in CC meteorites and that some rocks on the surface of Ryugu are also thought to be breccias \citep{Su19}. Imaging the Ryugu samples with a MASCam spare, as in \citet{O21b}, may provide further insight into the nature of the inclusions in the MASCOT rock. In the meantime, investigations of the Ryugu samples continue. There have already been preliminary reports of bright, millimeter-sized inclusions that are rich in carbonates \citep{L22,N22}, which may narrow the gap between the returned samples and the MASCam observations.

\section*{Acknowledgments}

This study was supported by JSPS International Planetary Network. The authors appreciate helpful comments by the editor and an anonymous reviewer.

ONC data can be retrieved from JAXA (\url{https://data.darts.isas.jaxa.jp/pub/hayabusa2/onc_bundle/}) as version~3c. The data analyzed in this paper is version~3d, which is scheduled to be archived later this year. NIRS3 data are archived in the PDS Small Bodies Node (\url{https://sbn.psi.edu/pds/resource/hayabusa2/nirs3.html}). The TIR data archive is \citet{Ok21}.

\bibliography{Ryugu}{}

\begin{thebibliography}{42}
\expandafter\ifx\csname natexlab\endcsname\relax\def\natexlab#1{#1}\fi

\bibitem[{{Domingue} {et~al.}(2021){Domingue}, {Kitazato}, {Matsuoka},
  {Yokota}, {Tatsumi}, {Iwata}, {Abe}, {Ohtake}, {Matsuura}, {Schr{\"o}der},
  {Vilas}, {Barucci}, {Brunetto}, {Takir}, {Le Corre}, \& {Moskovitz}}]{D21}
{Domingue}, D., {Kitazato}, K., {Matsuoka}, M., {et~al.} 2021, \psj, 2, 178

\bibitem[{{Galiano} {et~al.}(2020){Galiano}, {Palomba}, {D'Amore}, {Zinzi},
  {Dirri}, {Longobardo}, {Kitazato}, {Iwata}, {Matsuoka}, {Hiroi}, {Takir},
  {Nakamura}, {Abe}, {Ohtake}, {Matsuura}, {Watanabe}, {Yoshikawa}, {Saiki},
  {Tanaka}, {Okada}, {Yamamoto}, {Takei}, {Shirai}, {Hirata}, {Hirata},
  {Matsumoto}, \& {Tsuda}}]{G20}
{Galiano}, A., {Palomba}, E., {D'Amore}, M., {et~al.} 2020, \icarus, 351,
  113959

\bibitem[{{Grott} {et~al.}(2019){Grott}, {Knollenberg}, {Hamm}, {Ogawa},
  {Jaumann}, {Otto}, {Delbo}, {Michel}, {Biele}, {Neumann}, {Knapmeyer},
  {K{\"u}hrt}, {Senshu}, {Okada}, {Helbert}, {Maturilli}, {M{\"u}ller},
  {Hagermann}, {Sakatani}, {Tanaka}, {Arai}, {Mottola}, {Tachibana}, {Pelivan},
  {Drube}, {Vincent}, {Yano}, {Pilorget}, {Matz}, {Schmitz}, {Koncz},
  {Schr{\"o}der}, {Trauthan}, {Schlotterer}, {Krause}, {Ho}, \&
  {Moussi-Soffys}}]{G19}
{Grott}, M., {Knollenberg}, J., {Hamm}, M., {et~al.} 2019, Nature Astronomy, 3,
  971

\bibitem[{{Hamm} {et~al.}(2022){Hamm}, {Grott}, {Senshu}, {Knollenberg}, {de
  Wiljes}, {Hamilton}, {Scholten}, {Matz}, {Bates}, {Maturilli}, {Shimaki},
  {Sakatani}, {Neumann}, {Okada}, {Preusker}, {Elgner}, {Helbert}, {K{\"u}hrt},
  {Ho}, {Tanaka}, {Jaumann}, \& {Sugita}}]{H22}
{Hamm}, M., {Grott}, M., {Senshu}, H., {et~al.} 2022, Nature Communications,
  13, 364

\bibitem[{{Hirata} {et~al.}(2020){Hirata}, {Sugiyama}, {Hirata}, {Tanaka},
  {Nishikawa}, {Noguchi}, {Shimaki}, {Gaskell}, {Palmer}, {Matsumoto},
  {Senshu}, {Yamamoto}, {Murakami}, {Ishihara}, {Sugita}, {Morota}, {Honda},
  {Arakawa}, {Ogawa}, {Tsuda}, \& {Watanabe}}]{Hi20}
{Hirata}, N., {Sugiyama}, T., {Hirata}, N., {et~al.} 2020, in 51st Annual Lunar
  and Planetary Science Conference, Lunar and Planetary Science Conference,
  2015

\bibitem[{{Ho} {et~al.}(2017){Ho}, {Baturkin}, {Grimm}, {Grundmann}, {Hobbie},
  {Ksenik}, {Lange}, {Sasaki}, {Schlotterer}, {Talapina}, {Termtanasombat},
  {Wejmo}, {Witte}, {Wrasmann}, {W{\"u}bbels}, {R{\"o}{\ss}ler}, {Ziach},
  {Findlay}, {Biele}, {Krause}, {Ulamec}, {Lange}, {Mierheim}, {Lichtenheldt},
  {Maier}, {Reill}, {Sedlmayr}, {Bousquet}, {Bellion}, {Bompis},
  {Cenac-Morthe}, {Deleuze}, {Fredon}, {Jurado}, {Canalias}, {Jaumann},
  {Bibring}, {Glassmeier}, {Hercik}, {Grott}, {Celotti}, {Cordero},
  {Hendrikse}, \& {Okada}}]{H17}
{Ho}, T.-M., {Baturkin}, V., {Grimm}, C., {et~al.} 2017, \ssr, 208, 339

\bibitem[{{Ho} {et~al.}(2021){Ho}, {Jaumann}, {Bibring}, {Grott},
  {Gla{\ss}meier}, {Moussi}, {Krause}, {Auster}, {Baturkin}, {Biele},
  {Cordero}, {Cozzoni}, {Dudal}, {Fantinati}, {Grimm}, {Grundmann}, {Hamm},
  {Her{\v{c}}ik}, {Kayal}, {Knollenberg}, {K{\"u}chemann}, {Ksenik}, {Lange},
  {Lange}, {Lorda}, {Maibaum}, {Mimasu}, {Cenac-Morthe}, {Okada}, {Otto},
  {Pilorget}, {Reill}, {Saiki}, {Sasaki}, {Schlotterer}, {Schmitz},
  {Schr{\"o}der}, {Termtanasombat}, {Toth}, {Tsuda}, {Ulamec}, {Wolff},
  {Yoshimitsu}, {Ziach}, \& {MASCOT Team}}]{Ho20}
{Ho}, T.-M., {Jaumann}, R., {Bibring}, J.-P., {et~al.} 2021, \planss, 200,
  105200

\bibitem[{{Iwata} {et~al.}(2017){Iwata}, {Kitazato}, {Abe}, {Ohtake}, {Arai},
  {Arai}, {Hirata}, {Hiroi}, {Honda}, {Imae}, {Komatsu}, {Matsunaga},
  {Matsuoka}, {Matsuura}, {Nakamura}, {Nakato}, {Nakauchi}, {Osawa}, {Senshu},
  {Takagi}, {Tsumura}, {Takato}, {Watanabe}, {Barucci}, {Palomba}, \&
  {Ozaki}}]{I17}
{Iwata}, T., {Kitazato}, K., {Abe}, M., {et~al.} 2017, \ssr, 208, 317

\bibitem[{{Jaumann} {et~al.}(2019){Jaumann}, {Schmitz}, {Ho}, {Schr{\"o}der},
  {Otto}, {Stephan}, {Elgner}, {Krohn}, {Preusker}, {Scholten}, {Biele},
  {Ulamec}, {Krause}, {Sugita}, {Matz}, {Roatsch}, {Parekh}, {Mottola},
  {Grott}, {Michel}, {Trauthan}, {Koncz}, {Michaelis}, {Lange}, {Grundmann},
  {Maibaum}, {Sasaki}, {Wolff}, {Reill}, {Moussi-Soffys}, {Lorda}, {Neumann},
  {Vincent}, {Wagner}, {Bibring}, {Kameda}, {Yano}, {Watanabe}, {Yoshikawa},
  {Tsuda}, {Okada}, {Yoshimitsu}, {Mimasu}, {Saiki}, {Yabuta}, {Rauer},
  {Honda}, {Morota}, {Yokota}, \& {Kouyama}}]{J19}
{Jaumann}, R., {Schmitz}, N., {Ho}, T.~M., {et~al.} 2019, Science, 365, 817

\bibitem[{{Jaumann} {et~al.}(2017){Jaumann}, {Schmitz}, {Koncz}, {Michaelis},
  {Schroeder}, {Mottola}, {Trauthan}, {Hoffmann}, {Roatsch}, {Jobs},
  {Kachlicki}, {Pforte}, {Terzer}, {Tschentscher}, {Weisse}, {Mueller},
  {Perez-Prieto}, {Broll}, {Kruselburger}, {Ho}, {Biele}, {Ulamec}, {Krause},
  {Grott}, {Bibring}, {Watanabe}, {Sugita}, {Okada}, {Yoshikawa}, \&
  {Yabuta}}]{J17}
{Jaumann}, R., {Schmitz}, N., {Koncz}, A., {et~al.} 2017, \ssr, 208, 375

\bibitem[{{Kameda} {et~al.}(2017){Kameda}, {Suzuki}, {Takamatsu}, {Cho},
  {Yasuda}, {Yamada}, {Sawada}, {Honda}, {Morota}, {Honda}, {Sato}, {Okumura},
  {Shibasaki}, {Ikezawa}, \& {Sugita}}]{K17}
{Kameda}, S., {Suzuki}, H., {Takamatsu}, T., {et~al.} 2017, \ssr, 208, 17

\bibitem[{{Kitazato} {et~al.}(2019){Kitazato}, {Milliken}, {Iwata}, {Abe},
  {Ohtake}, {Matsuura}, {Arai}, {Nakauchi}, {Nakamura}, {Matsuoka}, {Senshu},
  {Hirata}, {Hiroi}, {Pilorget}, {Brunetto}, {Poulet}, {Riu}, {Bibring},
  {Takir}, {Domingue}, {Vilas}, {Barucci}, {Perna}, {Palomba}, {Galiano},
  {Tsumura}, {Osawa}, {Komatsu}, {Nakato}, {Arai}, {Takato}, {Matsunaga},
  {Takagi}, {Matsumoto}, {Kouyama}, {Yokota}, {Tatsumi}, {Sakatani},
  {Yamamoto}, {Okada}, {Sugita}, {Honda}, {Morota}, {Kameda}, {Sawada},
  {Honda}, {Yamada}, {Suzuki}, {Yoshioka}, {Hayakawa}, {Ogawa}, {Cho},
  {Shirai}, {Shimaki}, {Hirata}, {Yamaguchi}, {Ogawa}, {Terui}, {Yamaguchi},
  {Takei}, {Saiki}, {Nakazawa}, {Tanaka}, {Yoshikawa}, {Watanabe}, \&
  {Tsuda}}]{K19}
{Kitazato}, K., {Milliken}, R.~E., {Iwata}, T., {et~al.} 2019, Science, 364,
  272

\bibitem[{{Kouyama} {et~al.}(2021{\natexlab{a}}){Kouyama}, {Tatsumi}, {Honda},
  {Honda}, {Morota}, {Yokota}, {Kameda}, {Yamada}, {Suzuki}, {Sakatani},
  {Hayakawa}, {Cho}, {Matsuoka}, {Yoshioka}, {Sawada}, \& {Sugita}}]{K21b}
{Kouyama}, T., {Tatsumi}, E., {Honda}, C., {et~al.} 2021{\natexlab{a}}, arXiv
  e-prints, arXiv:2112.09404

\bibitem[{{Kouyama} {et~al.}(2021{\natexlab{b}}){Kouyama}, {Tatsumi}, {Yokota},
  {Yumoto}, {Yamada}, {Honda}, {Kameda}, {Suzuki}, {Sakatani}, {Hayakawa},
  {Morota}, {Matsuoka}, {Cho}, {Honda}, {Sawada}, {Yoshioka}, \&
  {Sugita}}]{K21}
{Kouyama}, T., {Tatsumi}, E., {Yokota}, Y., {et~al.} 2021{\natexlab{b}},
  \icarus, 360, 114353

\bibitem[{{Loizeau} {et~al.}(2022){Loizeau}, {Bibring}, {Brunetto}, {Pilorget},
  {Okada}, {Carter}, {Gondet}, {Hamm}, {Hatakeda}, {Langevin}, {Lantz}, {Le
  Pivert-Jolivet}, {Nakato}, {Riu}, {Usui}, {Yada}, \& {Yogata}}]{L22}
{Loizeau}, D., {Bibring}, J.~P., {Brunetto}, R., {et~al.} 2022, in LPI
  Contributions, Vol. 2678, LPI Contributions, 1495

\bibitem[{{Longobardo} {et~al.}(2019){Longobardo}, {Palomba}, {Galiano}, {De
  Sanctis}, {Ciarniello}, {Raponi}, {Tosi}, {Schr{\"o}der}, {Carrozzo},
  {Ammannito}, {Zambon}, {Stephan}, {Capria}, {Rognini}, {Raymond}, \&
  {Russell}}]{Lo19}
{Longobardo}, A., {Palomba}, E., {Galiano}, A., {et~al.} 2019, \icarus, 320, 97

\bibitem[{{Lorda} {et~al.}(2020){Lorda}, {Canalias}, {Martin}, {Garmier},
  {Moussi}, {Biele}, {Jaumann}, {Bibring}, {Grott}, {Auster}, {Ho}, {Krause},
  {Maibaum}, {Cozzoni}, {Ulamec}, {Wolff}, {Tsuda}, {Okada}, \& {Mimasu}}]{L20}
{Lorda}, L., {Canalias}, E., {Martin}, T., {et~al.} 2020, \planss, 194, 105086

\bibitem[{{Morota} {et~al.}(2020){Morota}, {Sugita}, {Cho}, {Kanamaru},
  {Tatsumi}, {Sakatani}, {Honda}, {Hirata}, {Kikuchi}, {Yamada}, {Yokota},
  {Kameda}, {Matsuoka}, {Sawada}, {Honda}, {Kouyama}, {Ogawa}, {Suzuki},
  {Yoshioka}, {Hayakawa}, {Hirata}, {Hirabayashi}, {Miyamoto}, {Michikami},
  {Hiroi}, {Hemmi}, {Barnouin}, {Ernst}, {Kitazato}, {Nakamura}, {Riu},
  {Senshu}, {Kobayashi}, {Sasaki}, {Komatsu}, {Tanabe}, {Fujii}, {Irie},
  {Suemitsu}, {Takaki}, {Sugimoto}, {Yumoto}, {Ishida}, {Kato}, {Moroi},
  {Domingue}, {Michel}, {Pilorget}, {Iwata}, {Abe}, {Ohtake}, {Nakauchi},
  {Tsumura}, {Yabuta}, {Ishihara}, {Noguchi}, {Matsumoto}, {Miura}, {Namiki},
  {Tachibana}, {Arakawa}, {Ikeda}, {Wada}, {Mizuno}, {Hirose}, {Hosoda},
  {Mori}, {Shimada}, {Soldini}, {Tsukizaki}, {Yano}, {Ozaki}, {Takeuchi},
  {Yamamoto}, {Okada}, {Shimaki}, {Shirai}, {Iijima}, {Noda}, {Kikuchi},
  {Yamaguchi}, {Ogawa}, {Ono}, {Mimasu}, {Yoshikawa}, {Takahashi}, {Takei},
  {Fujii}, {Nakazawa}, {Terui}, {Tanaka}, {Yoshikawa}, {Saiki}, {Watanabe}, \&
  {Tsuda}}]{M20}
{Morota}, T., {Sugita}, S., {Cho}, Y., {et~al.} 2020, Science, 368, 654

\bibitem[{{Nakato} {et~al.}(2022){Nakato}, {Yada}, {Yogata}, {Miyazaki},
  {Hatakeda}, {Kumagai}, {Nishimura}, {Hitomi}, {Soejima}, {Nagashima},
  {Bibring}, {Pilorget}, {Hamm}, {Brunetto}, {Riu}, {Lourit}, {Loizeau},
  {Pivert-Jolivet}, {Lequertier}, {Moussi-Soffys}, {Abe}, {Okada}, \&
  {Usui}}]{N22}
{Nakato}, A., {Yada}, T., {Yogata}, K., {et~al.} 2022, in LPI Contributions,
  Vol. 2678, LPI Contributions, 1810

\bibitem[{Okada {et~al.}(2020)Okada, Fukuhara, Tanaka, Taguchi, Arai, Senshu,
  Sakatani, Shimaki, Demura, Ogawa, Suko, Sekiguchi, Kouyama, Takita,
  Matsunaga, Imamura, Wada, Hasegawa, Helbert, M{\"u}ller, Hagermann, Biele,
  Grott, Hamm, Delbo, Hirata, Hirata, Yamamoto, Sugita, Namiki, Kitazato,
  Arakawa, Tachibana, Ikeda, Ishiguro, Wada, Honda, Honda, Ishihara, Matsumoto,
  Matsuoka, Michikami, Miura, Morota, Noda, Noguchi, Ogawa, Shirai, Tatsumi,
  Yabuta, Yokota, Yamada, Abe, Hayakawa, Iwata, Ozaki, Yano, Hosoda, Mori,
  Sawada, Shimada, Takeuchi, Tsukizaki, Fujii, Hirose, Kikuchi, Mimasu, Ogawa,
  Ono, Takahashi, Takei, Yamaguchi, Yoshikawa, Terui, Saiki, Nakazawa,
  Yoshikawa, Watanabe, \& Tsuda}]{O20}
Okada, T., Fukuhara, T., Tanaka, S., {et~al.} 2020, Nature, 579, 518

\bibitem[{{Okada} {et~al.}(2017){Okada}, {Fukuhara}, {Tanaka}, {Taguchi},
  {Imamura}, {Arai}, {Senshu}, {Ogawa}, {Demura}, {Kitazato}, {Nakamura},
  {Kouyama}, {Sekiguchi}, {Hasegawa}, {Matsunaga}, {Wada}, {Takita},
  {Sakatani}, {Horikawa}, {Endo}, {Helbert}, {M{\"u}ller}, \&
  {Hagermann}}]{O17}
{Okada}, T., {Fukuhara}, T., {Tanaka}, S., {et~al.} 2017, \ssr, 208, 255

\bibitem[{{Okada} {et~al.}(2021){Okada}, {Tanaka}, {Fukuhara}, {Arai},
  {Imamura}, {Ogawa}, {Kitazato}, {Kouyama}, {Sakatani}, {Shimaki},
  {Sekiguchi}, {Senshu}, {Takita}, {Taguchi}, {Demura}, {Nakamura}, {Hasegawa},
  {Matsunaga}, {Wada}, {Helbert}, {M{\"u}ller}, {Hagermann}, {Biele}, {Grott},
  {Hamm}, {Delbo}, {Murakami}, {Yamamoto}, {Crombie}, \& {Ishihara}}]{Ok21}
{Okada}, T., {Tanaka}, S., {Fukuhara}, T., {et~al.} 2021, PDS4 Hayabusa2 TIR
  Bundle, \url{https://doi.org/10.17597/isas.darts/hyb2-00300}

\bibitem[{{Otto} {et~al.}(2021){Otto}, {Schr{\"o}der}, {Scharf}, {Greshake},
  {Schmitz}, {Trauthan}, {Pieth}, {Stephan}, {Ho}, {Jaumann}, {Koncz},
  {Michalik}, \& {Yabuta}}]{O21b}
{Otto}, K.~A., {Schr{\"o}der}, S.~E., {Scharf}, H.~D., {et~al.} 2021, \psj, 2,
  188

\bibitem[{{Pilorget} {et~al.}(2021){Pilorget}, {Okada}, {Hamm}, {Brunetto},
  {Yada}, {Loizeau}, {Riu}, {Usui}, {Moussi-Soffys}, {Hatakeda}, {Nakato},
  {Yogata}, {Abe}, {Al{\'e}on-Toppani}, {Carter}, {Chaigneau}, {Crane},
  {Gondet}, {Kumagai}, {Langevin}, {Lantz}, {Le Pivert-Jolivet}, {Lequertier},
  {Lourit}, {Miyazaki}, {Nishimura}, {Poulet}, {Arakawa}, {Hirata}, {Kitazato},
  {Nakazawa}, {Namiki}, {Saiki}, {Sugita}, {Tachibana}, {Tanaka}, {Yoshikawa},
  {Tsuda}, {Watanabe}, \& {Bibring}}]{P21}
{Pilorget}, C., {Okada}, T., {Hamm}, V., {et~al.} 2021, Nature Astronomy, 6,
  221

\bibitem[{{Preusker} {et~al.}(2019){Preusker}, {Scholten}, {Elgner}, {Matz},
  {Kameda}, {Roatsch}, {Jaumann}, {Sugita}, {Honda}, {Morota}, {Tatsumi},
  {Cho}, {Yoshioka}, {Sawada}, {Yokota}, {Sakatani}, {Hayakawa}, {Matsuoka},
  {Yamada}, {Kouyama}, {Suzuki}, {Honda}, \& {Ogawa}}]{P19}
{Preusker}, F., {Scholten}, F., {Elgner}, S., {et~al.} 2019, \aap, 632, L4

\bibitem[{{Sakatani} {et~al.}(2021){Sakatani}, {Tanaka}, {Okada}, {Fukuhara},
  {Riu}, {Sugita}, {Honda}, {Morota}, {Kameda}, {Yokota}, {Tatsumi}, {Yumoto},
  {Hirata}, {Miura}, {Kouyama}, {Senshu}, {Shimaki}, {Arai}, {Takita},
  {Demura}, {Sekiguchi}, {M{\"u}ller}, {Hagermann}, {Biele}, {Grott}, {Hamm},
  {Delbo}, {Neumann}, {Taguchi}, {Ogawa}, {Matsunaga}, {Wada}, {Hasegawa},
  {Helbert}, {Hirata}, {Noguchi}, {Yamada}, {Suzuki}, {Honda}, {Ogawa},
  {Hayakawa}, {Yoshioka}, {Matsuoka}, {Cho}, {Sawada}, {Kitazato}, {Iwata},
  {Abe}, {Ohtake}, {Matsuura}, {Matsumoto}, {Noda}, {Ishihara}, {Yamamoto},
  {Higuchi}, {Namiki}, {Ono}, {Saiki}, {Imamura}, {Takagi}, {Yano}, {Shirai},
  {Okamoto}, {Nakazawa}, {Iijima}, {Arakawa}, {Wada}, {Kadono}, {Ishibashi},
  {Terui}, {Kikuchi}, {Yamaguchi}, {Ogawa}, {Mimasu}, {Yoshikawa}, {Takahashi},
  {Takei}, {Fujii}, {Takeuchi}, {Yamamoto}, {Hirose}, {Hosoda}, {Mori},
  {Shimada}, {Soldini}, {Tsukizaki}, {Ozaki}, {Tachibana}, {Ikeda}, {Ishiguro},
  {Yabuta}, {Yoshikawa}, {Watanabe}, \& {Tsuda}}]{Sa21}
{Sakatani}, N., {Tanaka}, S., {Okada}, T., {et~al.} 2021, Nature Astronomy, 5,
  766

\bibitem[{{Scholten} {et~al.}(2019{\natexlab{a}}){Scholten}, {Preusker},
  {Elgner}, {Matz}, {Jaumann}, {Biele}, {Hercik}, {Auster}, {Hamm}, {Grott},
  {Grimm}, {Ho}, {Koncz}, {Schmitz}, {Trauthan}, {Kameda}, {Sugita}, {Honda},
  {Morota}, {Tatsumi}, {Cho}, {Yoshioka}, {Sawada}, {Yokota}, {Sakatani},
  {Hayakawa}, {Matsuoka}, {Yamada}, {Kouyama}, {Suzuki}, {Honda}, \&
  {Ogawa}}]{S19b}
{Scholten}, F., {Preusker}, F., {Elgner}, S., {et~al.} 2019{\natexlab{a}},
  \aap, 632, L3

\bibitem[{{Scholten} {et~al.}(2019{\natexlab{b}}){Scholten}, {Preusker},
  {Elgner}, {Matz}, {Jaumann}, {Hamm}, {Schr{\"o}der}, {Koncz}, {Schmitz},
  {Trauthan}, {Grott}, {Biele}, {Ho}, {Kameda}, \& {Sugita}}]{S19a}
{Scholten}, F., {Preusker}, F., {Elgner}, S., {et~al.} 2019{\natexlab{b}},
  \aap, 632, L5

\bibitem[{{Schr{\"o}der} {et~al.}(2021){Schr{\"o}der}, {Otto}, {Scharf},
  {Matz}, {Schmitz}, {Scholten}, {Mottola}, {Trauthan}, {Koncz}, {Michaelis},
  {Jaumann}, {Ho}, {Yabuta}, \& {Sugita}}]{S20}
{Schr{\"o}der}, S., {Otto}, K., {Scharf}, H., {et~al.} 2021, \psj, 2, 58

\bibitem[{{Shimaki} {et~al.}(2020){Shimaki}, {Senshu}, {Sakatani}, {Okada},
  {Fukuhara}, {Tanaka}, {Taguchi}, {Arai}, {Demura}, {Ogawa}, {Suko},
  {Sekiguchi}, {Kouyama}, {Hasegawa}, {Takita}, {Matsunaga}, {Imamura}, {Wada},
  {Kitazato}, {Hirata}, {Hirata}, {Noguchi}, {Sugita}, {Kikuchi}, {Yamaguchi},
  {Ogawa}, {Ono}, {Mimasu}, {Yoshikawa}, {Takahashi}, {Takei}, {Fujii},
  {Takeuchi}, {Yamamoto}, {Yamada}, {Shirai}, {Iijima}, {Ogawa}, {Nakazawa},
  {Terui}, {Saiki}, {Yoshikawa}, {Tsuda}, \& {Watanabe}}]{Sh20}
{Shimaki}, Y., {Senshu}, H., {Sakatani}, N., {et~al.} 2020, \icarus, 348,
  113835

\bibitem[{{Sugimoto} {et~al.}(2021){Sugimoto}, {Tatsumi}, {Cho}, {Morota},
  {Honda}, {Kameda}, {Yokota}, {Yumoto}, {Aoki}, {DellaGiustina}, {Michikami},
  {Hiroi}, {Domingue}, {Michel}, {Schr{\"o}der}, {Nakamura}, {Yamada},
  {Sakatani}, {Kouyama}, {Honda}, {Hayakawa}, {Matsuoka}, {Suzuki}, {Yoshioka},
  {Ogawa}, {Sawada}, {Arakawa}, {Saiki}, {Imamura}, {Takagi}, {Yano}, {Shirai},
  {Okamoto}, {Tsuda}, {Nakazawa}, {Iijima}, \& {Sugita}}]{S21}
{Sugimoto}, C., {Tatsumi}, E., {Cho}, Y., {et~al.} 2021, \icarus, 369, 114591

\bibitem[{{Sugita} {et~al.}(2019){Sugita}, {Honda}, {Morota}, {Kameda},
  {Sawada}, {Tatsumi}, {Yamada}, {Honda}, {Yokota}, {Kouyama}, {Sakatani},
  {Ogawa}, {Suzuki}, {Okada}, {Namiki}, {Tanaka}, {Iijima}, {Yoshioka},
  {Hayakawa}, {Cho}, {Matsuoka}, {Hirata}, {Hirata}, {Miyamoto}, {Domingue},
  {Hirabayashi}, {Nakamura}, {Hiroi}, {Michikami}, {Michel}, {Ballouz},
  {Barnouin}, {Ernst}, {Schr{\"o}der}, {Kikuchi}, {Hemmi}, {Komatsu},
  {Fukuhara}, {Taguchi}, {Arai}, {Senshu}, {Demura}, {Ogawa}, {Shimaki},
  {Sekiguchi}, {M{\"u}ller}, {Hagermann}, {Mizuno}, {Noda}, {Matsumoto},
  {Yamada}, {Ishihara}, {Ikeda}, {Araki}, {Yamamoto}, {Abe}, {Yoshida},
  {Higuchi}, {Sasaki}, {Oshigami}, {Tsuruta}, {Asari}, {Tazawa}, {Shizugami},
  {Kimura}, {Otsubo}, {Yabuta}, {Hasegawa}, {Ishiguro}, {Tachibana}, {Palmer},
  {Gaskell}, {Le Corre}, {Jaumann}, {Otto}, {Schmitz}, {Abell}, {Barucci},
  {Zolensky}, {Vilas}, {Thuillet}, {Sugimoto}, {Takaki}, {Suzuki},
  {Kamiyoshihara}, {Okada}, {Nagata}, {Fujimoto}, {Yoshikawa}, {Yamamoto},
  {Shirai}, {Noguchi}, {Ogawa}, {Terui}, {Kikuchi}, {Yamaguchi}, {Oki},
  {Takao}, {Takeuchi}, {Ono}, {Mimasu}, {Yoshikawa}, {Takahashi}, {Takei},
  {Fujii}, {Hirose}, {Nakazawa}, {Hosoda}, {Mori}, {Shimada}, {Soldini},
  {Iwata}, {Abe}, {Yano}, {Tsukizaki}, {Ozaki}, {Nishiyama}, {Saiki},
  {Watanabe}, \& {Tsuda}}]{Su19}
{Sugita}, S., {Honda}, R., {Morota}, T., {et~al.} 2019, Science, 364, 252

\bibitem[{{Suzuki} {et~al.}(2018){Suzuki}, {Yamada}, {Kouyama}, {Tatsumi},
  {Kameda}, {Honda}, {Sawada}, {Ogawa}, {Morota}, {Honda}, {Sakatani},
  {Hayakawa}, {Yokota}, {Yamamoto}, \& {Sugita}}]{S18}
{Suzuki}, H., {Yamada}, M., {Kouyama}, T., {et~al.} 2018, \icarus, 300, 341

\bibitem[{{Tachibana} {et~al.}(2022){Tachibana}, {Sawada}, {Okazaki}, {Takano},
  {Sakamoto}, {Miura}, {Okamoto}, {Yano}, {Yamanouchi}, {Michel}, {Zhang},
  {Schwartz}, {Thuillet}, {Yurimoto}, {Nakamura}, {Noguchi}, {Yabuta},
  {Naraoka}, {Tsuchiyama}, {Imae}, {Kurosawa}, {Nakamura}, {Ogawa}, {Sugita},
  {Morota}, {Honda}, {Kameda}, {Tatsumi}, {Cho}, {Yoshioka}, {Yokota},
  {Hayakawa}, {Matsuoka}, {Sakatani}, {Yamada}, {Kouyama}, {Suzuki}, {Honda},
  {Yoshimitsu}, {Kubota}, {Demura}, {Yada}, {Nishimura}, {Yogata}, {Nakato},
  {Yoshitake}, {Suzuki}, {Furuya}, {Hatakeda}, {Miyazaki}, {Kumagai}, {Okada},
  {Abe}, {Usui}, {Ireland}, {Fujimoto}, {Yamada}, {Arakawa}, {Connolly},
  {Fujii}, {Hasegawa}, {Hirata}, {Hirata}, {Hirose}, {Hosoda}, {Iijima},
  {Ikeda}, {Ishiguro}, {Ishihara}, {Iwata}, {Kikuchi}, {Kitazato}, {Lauretta},
  {Libourel}, {Marty}, {Matsumoto}, {Michikami}, {Mimasu}, {Miura}, {Mori},
  {Nakamura-Messenger}, {Namiki}, {Nguyen}, {Nittler}, {Noda}, {Noguchi},
  {Ogawa}, {Ono}, {Ozaki}, {Senshu}, {Shimada}, {Shimaki}, {Shirai}, {Soldini},
  {Takahashi}, {Takei}, {Takeuchi}, {Tsukizaki}, {Wada}, {Yamamoto},
  {Yoshikawa}, {Yumoto}, {Zolensky}, {Nakazawa}, {Terui}, {Tanaka}, {Saiki},
  {Yoshikawa}, {Watanabe}, \& {Tsuda}}]{T22}
{Tachibana}, S., {Sawada}, H., {Okazaki}, R., {et~al.} 2022, Science, 375, 1011

\bibitem[{{Takita} {et~al.}(2017){Takita}, {Senshu}, \& {Tanaka}}]{T17}
{Takita}, J., {Senshu}, H., \& {Tanaka}, S. 2017, \ssr, 208, 287

\bibitem[{{Tatsumi} {et~al.}(2020{\natexlab{a}}){Tatsumi}, {Domingue},
  {Schr{\"o}der}, {Yokota}, {Kuroda}, {Ishiguro}, {Hasegawa}, {Hiroi}, {Honda},
  {Hemmi}, {Le Corre}, {Sakatani}, {Morota}, {Yamada}, {Kameda}, {Koyama},
  {Suzuki}, {Cho}, {Yoshioka}, {Matsuoka}, {Honda}, {Hayakawa}, {Hirata},
  {Hirata}, {Yamamoto}, {Vilas}, {Takato}, {Yoshikawa}, {Abe}, \&
  {Sugita}}]{TD20}
{Tatsumi}, E., {Domingue}, D., {Schr{\"o}der}, S., {et~al.} 2020{\natexlab{a}},
  \aap, 639, A83

\bibitem[{{Tatsumi} {et~al.}(2019){Tatsumi}, {Kouyama}, {Suzuki}, {Yamada},
  {Sakatani}, {Kameda}, {Yokota}, {Honda}, {Morota}, {Moroi}, {Tanabe},
  {Kamiyoshihara}, {Ishida}, {Yoshioka}, {Sato}, {Honda}, {Hayakawa},
  {Kitazato}, {Sawada}, \& {Sugita}}]{T19}
{Tatsumi}, E., {Kouyama}, T., {Suzuki}, H., {et~al.} 2019, \icarus, 325, 153

\bibitem[{{Tatsumi} {et~al.}(2021){Tatsumi}, {Sakatani}, {Riu}, {Matsuoka},
  {Honda}, {Morota}, {Kameda}, {Nakamura}, {Zolensky}, {Brunetto}, {Hiroi},
  {Sasaki}, {Watanabe}, {Tanaka}, {Takita}, {Pilorget}, {de Le{\'o}n},
  {Popescu}, {Rizos}, {Licandro}, {Palomba}, {Domingue}, {Vilas}, {Campins},
  {Cho}, {Yoshioka}, {Sawada}, {Yokota}, {Hayakawa}, {Yamada}, {Kouyama},
  {Suzuki}, {Honda}, {Ogawa}, {Kitazato}, {Hirata}, {Hirata}, {Tsuda},
  {Yoshikawa}, {Saiki}, {Terui}, {Nakazawa}, {Takei}, {Takeuchi}, {Yamamoto},
  {Okada}, {Shimaki}, {Shirai}, \& {Sugita}}]{T21}
{Tatsumi}, E., {Sakatani}, N., {Riu}, L., {et~al.} 2021, Nature Communications,
  12, 5837

\bibitem[{{Tatsumi} {et~al.}(2020{\natexlab{b}}){Tatsumi}, {Sugimoto}, {Riu},
  {Sugita}, {Nakamura}, {Hiroi}, {Morota}, {Popescu}, {Michikami}, {Kitazato},
  {Matsuoka}, {Kameda}, {Honda}, {Yamada}, {Sakatani}, {Kouyama}, {Yokota},
  {Honda}, {Suzuki}, {Cho}, {Ogawa}, {Hayakawa}, {Sawada}, {Yoshioka},
  {Pilorget}, {Ishida}, {Domingue}, {Hirata}, {Sasaki}, {de Le{\'o}n},
  {Barucci}, {Michel}, {Suemitsu}, {Saiki}, {Tanaka}, {Terui}, {Nakazawa},
  {Kikuchi}, {Yamaguchi}, {Ogawa}, {Ono}, {Mimasu}, {Yoshikawa}, {Takahashi},
  {Takei}, {Fujii}, {Yamamoto}, {Okada}, {Hirose}, {Hosoda}, {Mori}, {Shimada},
  {Soldini}, {Tsukizaki}, {Mizuno}, {Iwata}, {Yano}, {Ozaki}, {Abe}, {Ohtake},
  {Namiki}, {Tachibana}, {Arakawa}, {Ikeda}, {Ishiguro}, {Wada}, {Yabuta},
  {Takeuchi}, {Shimaki}, {Shirai}, {Hirata}, {Iijima}, {Tsuda}, {Watanabe}, \&
  {Yoshikawa}}]{TS20}
{Tatsumi}, E., {Sugimoto}, C., {Riu}, L., {et~al.} 2020{\natexlab{b}}, Nature
  Astronomy

\bibitem[{{Yada} {et~al.}(2021){Yada}, {Abe}, {Okada}, {Nakato}, {Yogata},
  {Miyazaki}, {Hatakeda}, {Kumagai}, {Nishimura}, {Hitomi}, {Soejima},
  {Yoshitake}, {Iwamae}, {Furuya}, {Uesugi}, {Karouji}, {Usui}, {Hayashi},
  {Yamamoto}, {Fukai}, {Sugita}, {Cho}, {Yumoto}, {Yabe}, {Bibring},
  {Pilorget}, {Hamm}, {Brunetto}, {Riu}, {Lourit}, {Loizeau}, {Lequertier},
  {Moussi-Soffys}, {Tachibana}, {Sawada}, {Okazaki}, {Takano}, {Sakamoto},
  {Miura}, {Yano}, {Ireland}, {Yamada}, {Fujimoto}, {Kitazato}, {Namiki},
  {Arakawa}, {Hirata}, {Yurimoto}, {Nakamura}, {Noguchi}, {Yabuta}, {Naraoka},
  {Ito}, {Nakamura}, {Uesugi}, {Kobayashi}, {Michikami}, {Kikuchi}, {Hirata},
  {Ishihara}, {Matsumoto}, {Noda}, {Noguchi}, {Shimaki}, {Shirai}, {Ogawa},
  {Wada}, {Senshu}, {Yamamoto}, {Morota}, {Honda}, {Honda}, {Yokota},
  {Matsuoka}, {Sakatani}, {Tatsumi}, {Miura}, {Yamada}, {Fujii}, {Hirose},
  {Hosoda}, {Ikeda}, {Iwata}, {Kikuchi}, {Mimasu}, {Mori}, {Ogawa}, {Ono},
  {Shimada}, {Soldini}, {Takahashi}, {Takei}, {Takeuchi}, {Tsukizaki},
  {Yoshikawa}, {Terui}, {Nakazawa}, {Tanaka}, {Saiki}, {Yoshikawa}, {Watanabe},
  \& {Tsuda}}]{Ya21}
{Yada}, T., {Abe}, M., {Okada}, T., {et~al.} 2021, Nature Astronomy, 6, 214

\bibitem[{{Yokota} {et~al.}(2021){Yokota}, {Honda}, {Tatsumi}, {Domingue},
  {Schr{\"o}der}, {Matsuoka}, {Morota}, {Sakatani}, {Kameda}, {Kouyama},
  {Yamada}, {Honda}, {Hayakawa}, {Cho}, {Michikami}, {Suzuki}, {Yoshioka},
  {Sawada}, {Ogawa}, {Yumoto}, \& {Sugita}}]{Yo21}
{Yokota}, Y., {Honda}, R., {Tatsumi}, E., {et~al.} 2021, \psj, 2, 177

\bibitem[{{Yumoto} {et~al.}(2022){Yumoto}, {Cho}, {Yabe}, {Mori}, {Ogura},
  {Miyazaki}, {Yada}, {Hatakeda}, {Yogata}, {Abe}, {Okada}, {Nishimura},
  {Usui}, \& {Sugita}}]{Y22}
{Yumoto}, K., {Cho}, Y., {Yabe}, Y., {et~al.} 2022, in LPI Contributions, Vol.
  2678, LPI Contributions, 1326

\end{thebibliography}
\bibliographystyle{aa} 



\begin{appendix}

\section{Spectra for other ONC image sets}

\begin{figure*}
        \centering
        \includegraphics[width=.49\textwidth,angle=0]{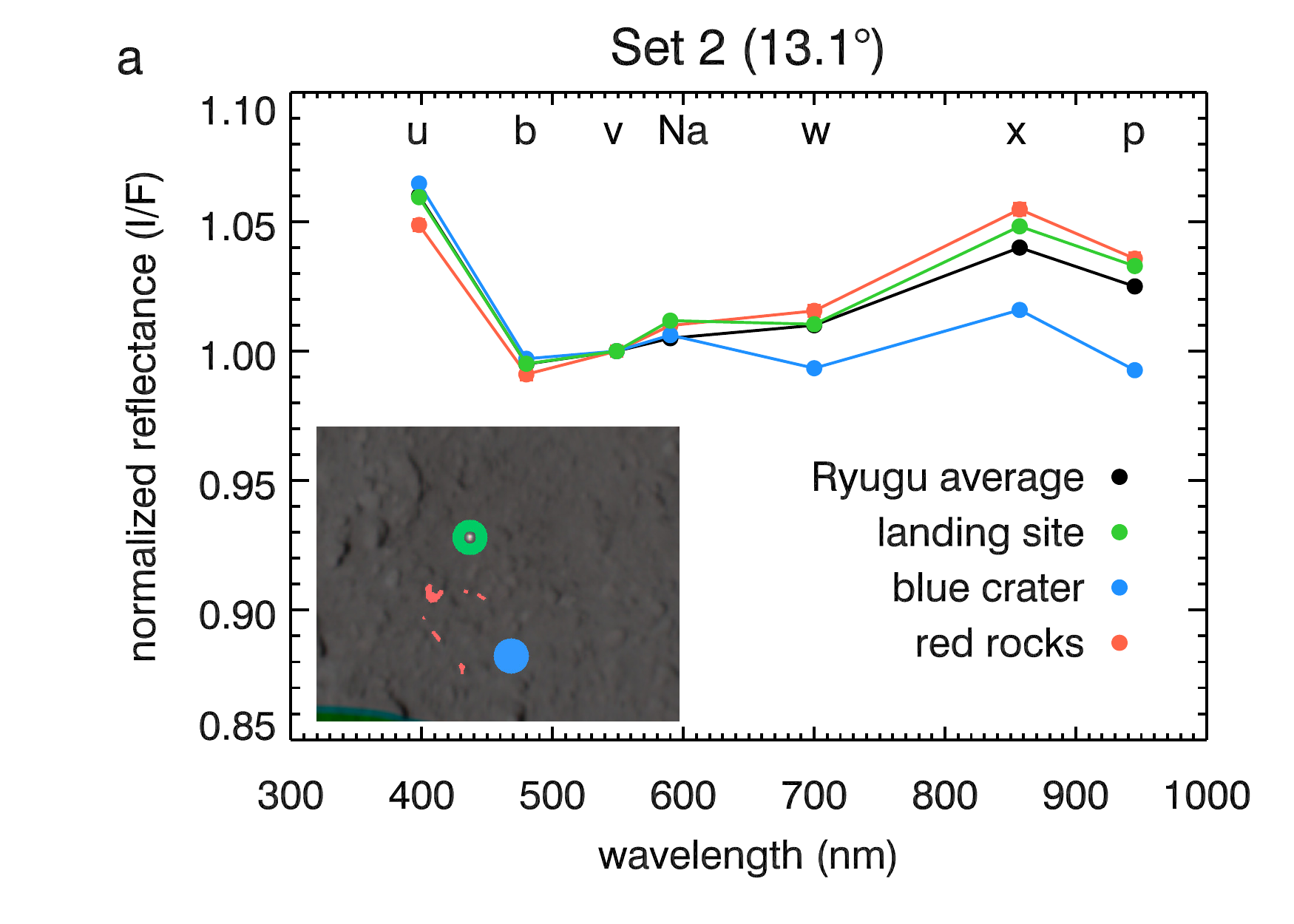}
        \includegraphics[width=.49\textwidth,angle=0]{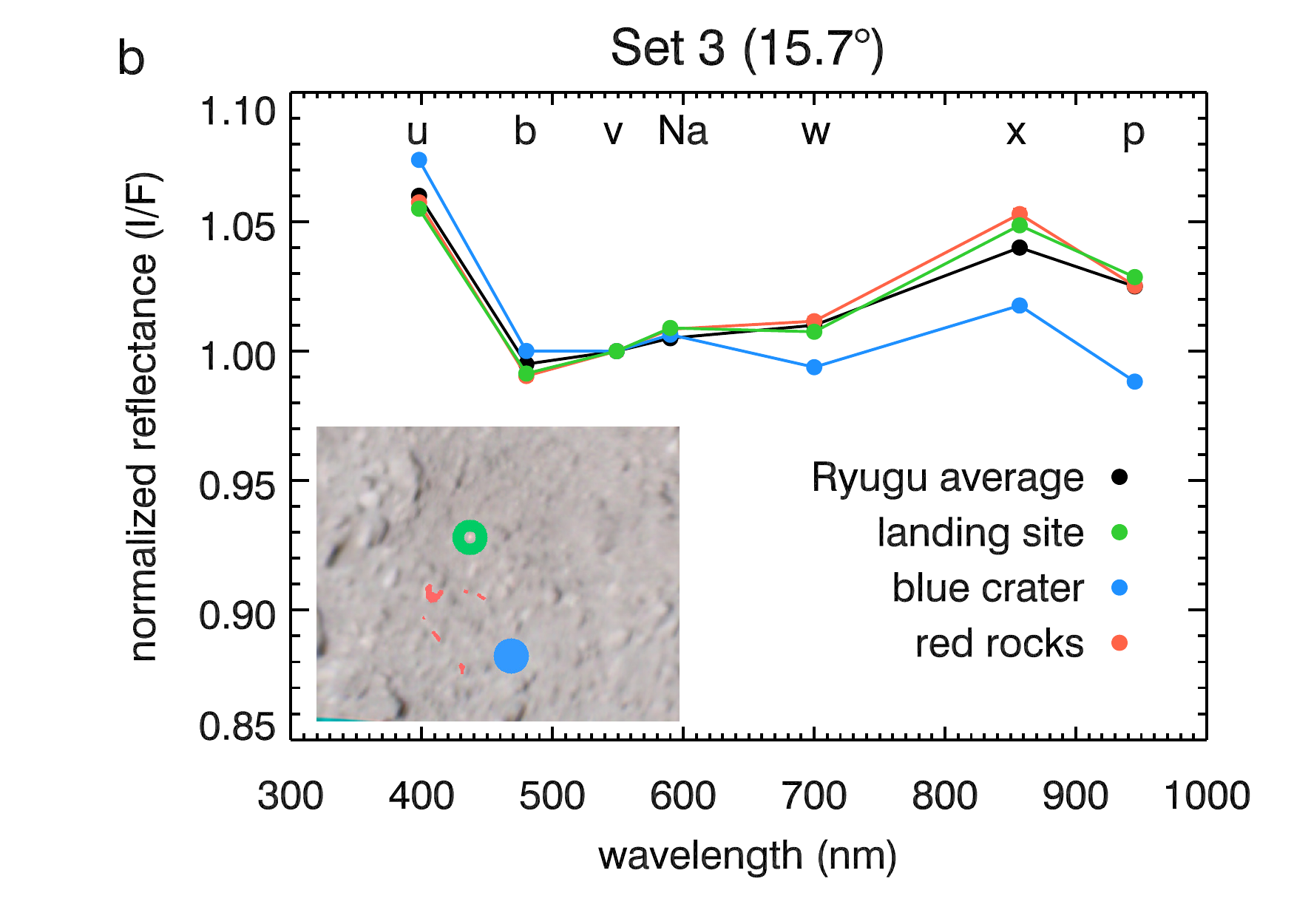}
        \includegraphics[width=.49\textwidth,angle=0]{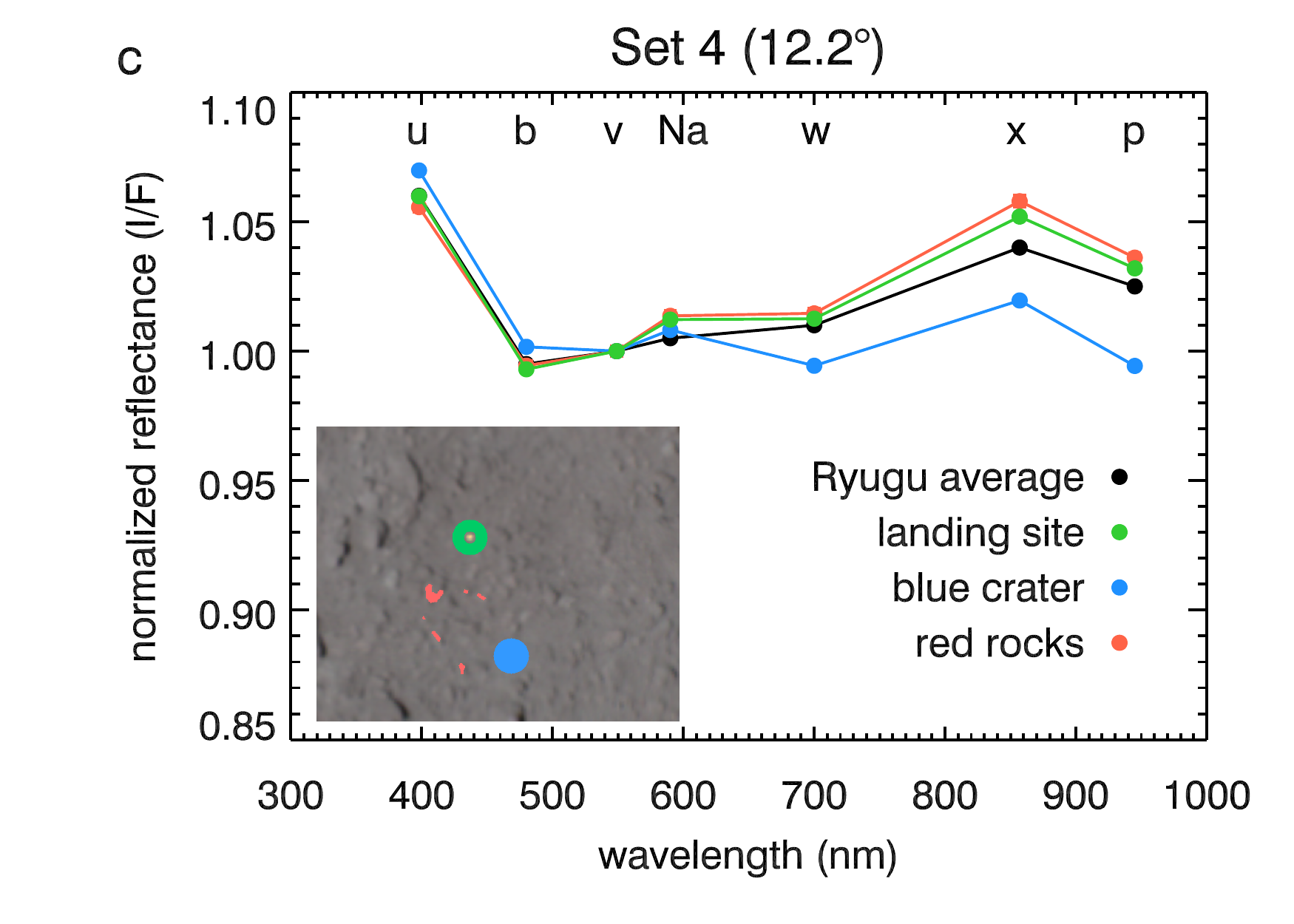}
        \includegraphics[width=.49\textwidth,angle=0]{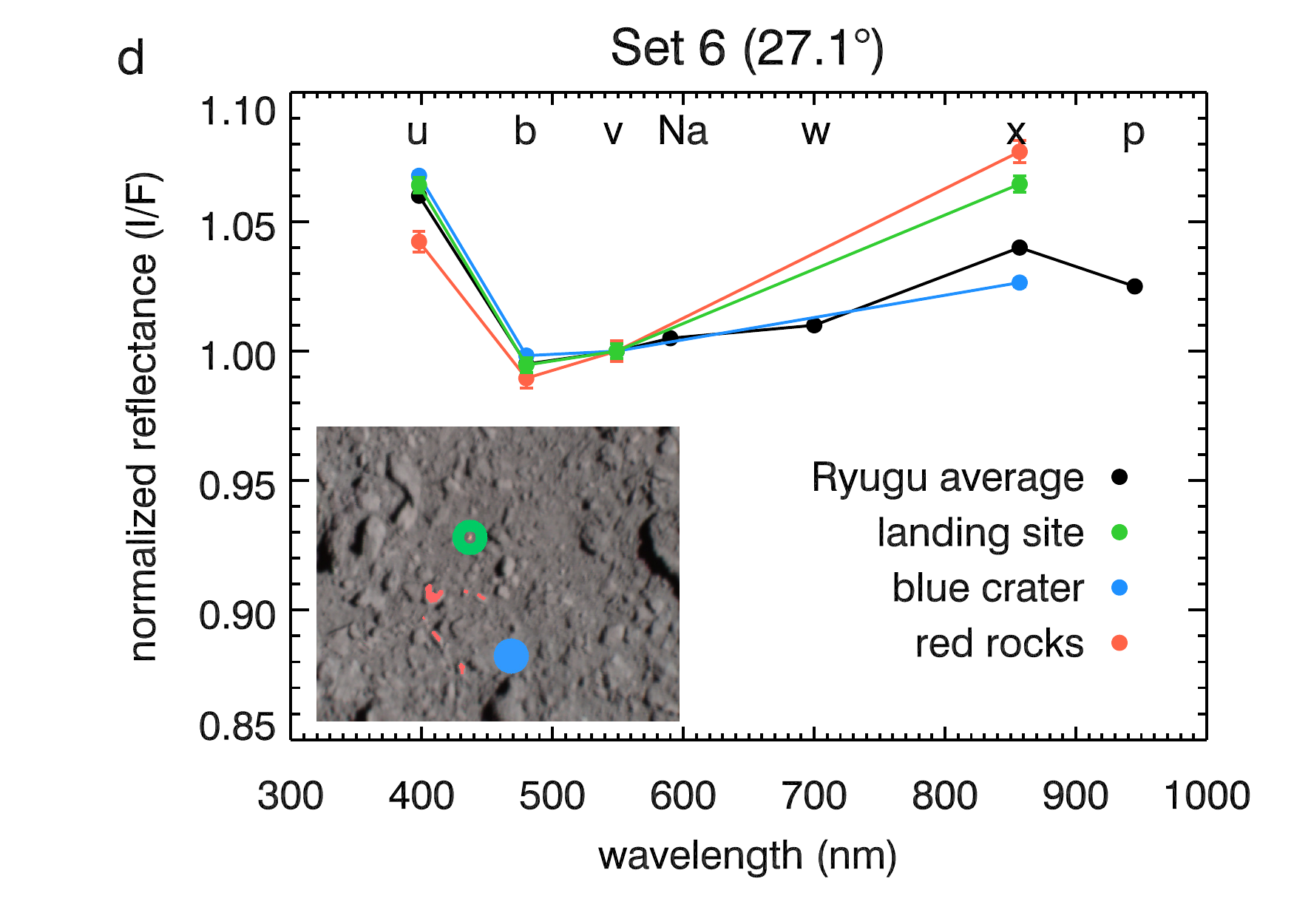}
        \caption{Spectra of different terrain units in the MASCOT landing area for ONC image sets~2 ({\bf a}), 3 ({\bf b}), 4 ({\bf c}), and 6 ({\bf d}), with the average phase angle indicated in the plot title. Error bars indicate the standard error of the mean and are generally smaller than the plot symbols. The insets show color composites of the respective sets, including the locations of the pixels that were averaged for each unit in the legend. The ``landing site'' unit excludes the lander itself. Spectra are normalized at 550~nm, and letters correspond to ONC filter names. The Ryugu average spectrum is that for the $19.5^\circ$ phase angle in Fig.~4 of \citet{TD20}.}
        \label{fig:appendix}
\end{figure*}

We analyzed a total of six ONC image sets, as listed in Table~\ref{tab:image_sets}. In Fig.~\ref{fig:spectral} we only show spectra for image sets 1 and 5, as their average phase angle is closest to that of the Ryugu spectrum we used for reference. Figure~\ref{fig:appendix} includes spectral plots for the other four image sets (2, 3, 4, and 6) to demonstrate the consistency of the results.

\end{appendix}

\end{document}